\documentclass[12pt,a4paper]{iopart}

\usepackage{rotating,graphicx}
\usepackage{amsfonts}%
\usepackage{caption}
\usepackage[utf8]{inputenc}
\usepackage[english]{babel} %francais, polish, spanish, ...
\usepackage[T1]{fontenc}
\usepackage{color}
\usepackage[colorlinks]{hyperref}
\usepackage{setspace}
\usepackage{layouts}
\begin{document}

\title[Characterization of the E$_r$ and edge velocity shear in W7-X]{Characterization of the radial electric field and edge velocity shear in Wendelstein 7-X}
\author{D. Carralero$^1$, T. Estrada$^1$,T. Windisch$^2$, J. L. Velasco$^1$, J. A. Alonso$^1$, M. Beurskens$^2$, S. Bozhenkov$^2$, H. Damm$^2$, G. Fuchert$^2$, Y. Gao$^2$, M. Jakubowski$^2$, H. Nieman$^2$, N. Pablant$^3$, E. Pasch$^2$, G. Weir$^2$, and the Wendelstein 7-X team.}
\address{$^1$ Laboratorio Nacional de Fusión. CIEMAT, 28040 Madrid, Spain.}
\address{$^2$ Max-Planck-Institut für Plasmaphysik, D-17491 Greifswald, Germany.}
\address{$^3$ Princeton Plasma Phys Lab, 100 Stellarator Rd, Princeton, NJ 08540 USA} 
\ead{daniel.carralero@ciemat.es}

\begin{abstract}

In this work we present the first measurements obtained by the V-band Doppler reflectometer during the second operation phase of Wendelstein 7-X to discuss the influence in the velocity shear layer and the radial electric field, E$_r$, of several plasma parameters such as magnetic configuration, rotational transform or degree of detachment. In the first place, we carry out a systematic characterization of the turbulence rotation velocity profile in order to describe the influence of density and heating power on E$_r$ under the four most frequent magnetic configurations. The $|$E$_r|$ value in the edge is found to increase with configurations featuring higher $\iota$, although this does not apply for the high mirror configuration, KJM. As well, the E$_r$ value in the SOL and the velocity shear near the separatrix are found to display a clear dependence on heating power and density for all configurations. For a number of relevant cases, these results are assessed by comparing them to neoclassical predictions obtained from the codes DKES and KNOSOS, finding generally good agreement with experimental results. Finally, the evolution of E$_r$ at the edge is evaluated throughout the island-divertor detachment regime achieved for the first time in the 2018 campaign. After detachment, $|$E$_r|$ is reduced both at the SOL and edge, and the plasma column shrinks, with the shear layer seemingly moving radially inwards from the separatrix.

\end{abstract}

\maketitle

\section{Introduction}\label{intro}

Wendelstein 7-X (W7-X) is a novel kind of optimized stellarator in which the concept of quasi-isodinamicity is used to reduce drift orbit losses caused by the three-dimensional magnetic field inhomogeneity combined with collisions, thus addressing the traditional low confinement of stellarators at low collisionalities \cite{Grieger92}. This device, which began operation in 2016 \cite{ref1}, has conducted its third experimental campaign (OP1.2b) between the months of August and October of 2018 \cite{ref1b}. While it is generally accepted that turbulence dominates transport in the edge of stellarators -where neoclassical predictions systematically underestimate radial fluxes \cite{ref3b}-, in this campaign it was also shown to be relevant in the core of W7-X \cite{ref3c}, where it has been proposed as the fundamental mechanism behind phenomena such as the lack of impurity accumulation \cite{ref3d} or the upper limit observed in the ion temperature, T$_i$, at the core in gas-fueled discharges \cite{ref3c}. These results indicate that finding mechanisms to regulate turbulence and turbulence-related transport can be determinant for the performance of future reactors based on the helias concept. One such mechanism is the  interplay between E$_r$ shear and edge turbulence: indeed, the suppression of turbulence associated to the presence of strong velocity shear layers has been related in the literature to the formation of transport barriers \cite{ref8}. For example, in the stellarator Wendelstein 7-AS (W7-AS) strong E$_r$ shear values were a common ingredient in several of the improved confinement regimes: it was found that both the Optimum Confinement mode \cite{ref8b} and the H-mode \cite{ref8c}. Already in W7-X, suppression of broadband turbulence in the vicinity of the edge velocity shear was reported in the limited density conditions achieved in the 2017 campaign \cite{AKF2019}. \\

Besides demonstrating its transport-related optimization, another goal of W7-X is to demonstrate the feasibility of stationary operation using an island divertor geometry \cite{Strumberger96} to handle heat and particle fluxes coming from the plasma onto the wall. In this regard, one of the main highlights of the 2018 campaign has been the achievement of a stable heat and particle detachment, which could be maintained during almost 30 seconds without significant reduction in confinement properties, while keeping high neutral pressures at the divertor and low impurity concentrations in the main plasma \cite{Jakub20}. Under this regime, most of the power leaving the plasma is dissipated by radiation before reaching the walls, with radiated fraction of power, $f_{rad} = P_{rad}/P_{ECRH}$, approaching unity as the detachment is complete. E$_r$ shear and edge turbulence are also linked to Scrape-off Layer (SOL) transport and divertor regime as they are often related in the literature to seeding/ejection rate for filamentary structures responsible for convective transport in the SOL \cite{ref8d}. In this sense, enhanced perpendicular transport regimes have been recently reported in tokamaks \cite{Carralero14} when the divertor approaches the roll-over. The evaluation of E$_r$ near the separatrix during detachment and its potential impact on turbulence is then  interesting, as it can shed light on these phenomena and their potential relevance for stellarators featuring an island divertor. \\

A Doppler reflectometer (DR) launches a microwave beam onto the plasma with an oblique incidence  \cite{Hirsch99} and measures the Bragg backscattering on plasma density fluctuations at the cutoff layer \cite{Hirsch01}. This diagnostic is particularly well suited to analyze the relation between electric fields and turbulence, as it can measure both simultaneously (see, eg. \cite{Estrada11, Conway11}): by evaluating the Doppler shift of the backscattered beam, the perpendicular velocity of fluctuations is measured in the laboratory frame thus allowing, under some assumptions, for an estimation of the local E$_r$. As well, the  backscattered power is related to the amplitude of the density fluctuations (if turbulence is not too intense, power can be considered proportional to the square of fluctuation amplitude, $S\propto \delta n_{rms}^2$ \cite{Gusakov04, Blanco08}), thus providing a $k_\perp$ selective monitor of turbulence. Thanks to another accomplishment of the 2018 campaign -the routinary use of boronization to condition the wall-, substantially higher densities were typically achieved in H plasmas than in preceding operation \cite{ref3}. In particular, density profiles compatible with operation for the V-band DR became commonplace during the campaign, thus allowing for the evaluation of rotation, radial electric field, E$_r$, and fluctuation amplitude under a wide number of magnetic configurations and operational scenarios.\\

In this work we set out to lay the foundation for this kind of studies at W7-X by reporting a systematic characterization of the measurements of radial electric field, E$_r$, at the edge and its variation in the vicinity of the separatrix, $\Delta$E$_r$, (which will be used as a proxy for the velocity shear) in order to describe the influence of density and heating power under the four most frequent magnetic configurations. As frequently done in stellarator literature (see eg. \cite{Ida05}, \cite{Velasco13} or \cite{Hirsch08} and references therein) observed fields have been compared to  neoclassical  simulations. Then, the onset of detachment in the standard configuration has been taken as a study case: the evolution of the E$_r$ profile is evaluated during the whole process and results are discussed to some detail. The relation between E$_r$ profiles, $\Delta$E$_r$ and turbulence amplitude is considered a second part of this study and has been left for future work. The paper is organized as follows: first we present the diagnostics and methodology in section \ref{Met} and describe the parametric studies and discuss their main results in section \ref{param}. Next, we compare those results with neoclassical simulations in section \ref{neo}. In section \ref{SL}, we take on the evolution of the E$_r$ profile during detachment. Finally, we discuss the results in section \ref{Disc} and outline our main conclusions in section \ref{Conc}.

\section{Methodology}\label{Met}

The diagnostic layout of Wendelstein 7-X features several reflectometer systems \cite{ref2}, including the main diagnostic used for this work: one of the two monostatic frequency hopping Doppler reflectometers (DR) installed at the elliptical toroidal section, below the equatorial plane. This system uses frequencies in the V-band ($50-75$ GHz) with O-mode polarization to probe the plasma with a fixed injection angle \cite{ref5}. As a result, the perpendicular wave numbers associated to the backscattering remain relatively unchanged, $k_{\perp} \simeq 7-10$ cm$^{-1}$. Along  this work, $E\times B$ bulk plasma flows are always considered dominant with respect to turbulence phase velocity ($v_E = |\mathbf{E}\times \mathbf{B}|/B^2 \gg v_{ph}$), thereby allowing the direct calculation of E$_r$ from the local value of the magnetic field and the turbulence rotation measurements provided by the DR \cite{Hirsch01}. The validity of this frequent hypothesis has been proven in a range of scenarios both in tokamaks, such as ASDEX-Upgrade \cite{Nold10} and stellarators, such as TJ-II \cite{Estrada09} or W7-AS \cite{Hirsch01}, although it is difficult to assess for any given measurement. In the case of W7-X, it has been confirmed (at least for low densities) in the SOL by comparison to E$_r$ profiles measured by Langmuir probes \cite{AKF2019}. Regarding the plasma edge, this hypothesis is generally confirmed in the cases under study by the good agreement with neoclassical calculations, although a more detailed discussion on the validity of this approach can be found elsewhere \cite{ref5}. Under typical settings, the reflectometer emitted series of $10$ ms long, $1$ GHz frequency steps from $50$ to $75$ GHz providing one radial profile of E$_r$ every $250$ ms. The emitted beam propagates through the plasma, being backscattered at the cutoff layer, in which its refractive index, $N$, reaches its minimum value. 
%(at this point, the perpendicular component of the refractive index becomes $N_\perp = 0$).
 Under O-mode polarization, $N$ is a function of beam frequency, $\omega$, and the plasma frequency,
\begin{equation}
N_O^2=\biggl(1-\frac{\omega_p^2}{\omega^2}\biggr), \label{eq0}
\end{equation}
where  $\omega_p=(n_ee^2/\epsilon_0m_e)^{1/2}$ and $n_e$, $e$ and $m_e$ stand for the electron density, charge and mass. Therefore, its position depends only on the local density and the trajectory of the beam. Under the slab approximation and assuming small refraction, the density at the cutoff layer, $n_{co}$, can be simply calculated for a given beam frequency as a function of the launching angle of the beam, $\theta_0$, by setting the condition $N_{co}=\sin{\theta_0}$. In a more realistic setting, ray tracing codes are required in order to calculate the precise value of $N_{co}$ (In this case, we used the Travis code \cite{Marush14} to perform such calculations). In either case, a measurement of the density profile from an independent diagnostic is required to determine the radial position of the cutoff layer. Along this work, Thomson scattering (TS) measurements \cite{Pasch16} were used for this. Since this diagnostic is installed in a different toroidal section, the density profile is expressed using the radial magnetic coordinate $\rho \equiv (\Psi/\Psi_{LCFS})^{1/2}$, where $\Psi$ is the toroidal flux through the magnetic surface and LCFS denotes the last-closed flux-surface. The $\rho$ position of TS data points is calculated using VMEC equilibria corresponding to the average $\left\langle \beta\right\rangle$ of each discharge. \\

\begin{figure}
	\centering
	\includegraphics[width=0.6\linewidth]{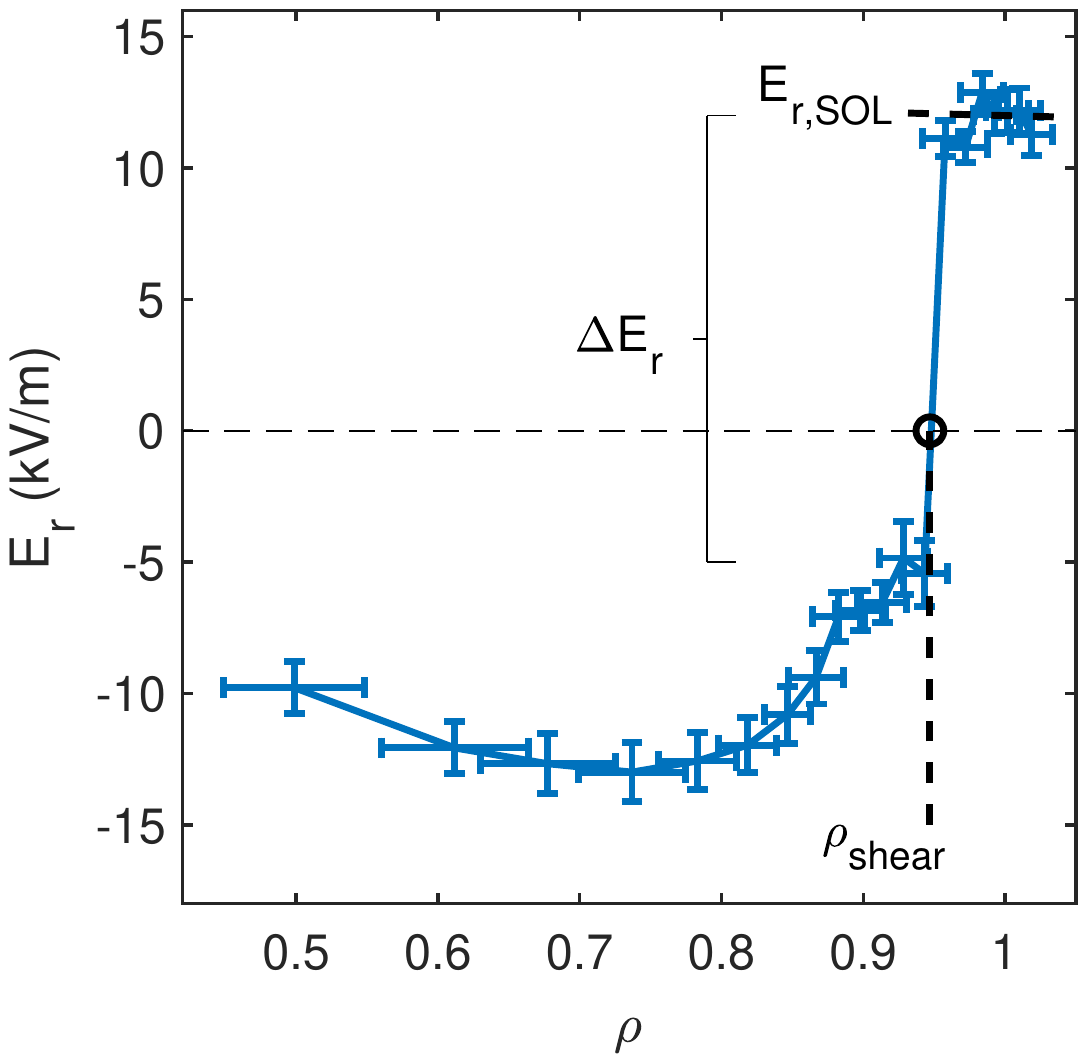}
	\caption{\textit{Example of a typical E$_r$ profile (shot $180920017$, t= $4000$ ms). Radial electric field variation over the shear, $\Delta$E$_r$, average E$_r$ at the SOL, E$_{r,SOL}$, and radial position of the shear, $\rho_{shear}$, are graphically defined.}}
	\label{fig0}
\end{figure}

Under most scenarios, the cutoff density for $50$ GHz is found in the SOL, while for $75$ GHz is typically inside the last closed flux surface (LCFS). As a result, the transition between open and closed field lines happens at radial positions covered by the radial range of the DR, and is reflected in the data as a sign reversal that typically appears around the LCFS when the negative ion root E$_r$ -found for typical W7-X operational conditions in the edge of the confined region \cite{ref4e,ref4f}- transits to the electron transport-dominated positive electric field at open field lines, determined at the wall by the cross-field gradient of the electron temperature, $\nabla_\bot$T$_e$. Since T$_e$ decreases when moving away from the strike point, this typically results in positive electric fields in the SOL. This kind of transition is commonplace in the literature, and experiments carried out in LHD have even related the magnitude of E$_r$ in the SOL to the local value of  $\nabla_\bot$T$_e$ \cite{Suzuki16} . When probing the boundary between the previously described regions, characterized by a poloidal velocity shear caused by the steep radial gradient of E$_r$, often two Doppler peaks of different sign are detected by the DR. As explained in \cite{Happel10}, this is an effect of the finite spectral resolution of the diagnostic, which measures both regions at the same time. As proposed in that reference, the peak with the highest spectral power is taken to determine the local E$_r$. A typical example of the edge radial electric field is displayed in figure \ref{fig0}. Vertical error bars reflect the uncertainty on $k_\perp$ and the fit error on the determination of the backscattered beam Doppler frequency. Horizontal error bars are calculated using Travis as the dispersion of the cutoff positions of a bundle of rays oriented as the one launched by the DR, but originated at several points of a ring surrounding the central one. These rays represent the envelope of the microwave beam where the amplitude dropped to $1/e$ of its central value and are used to simulate the finite extension of the backscattering region. This way, they provide the uncertainty on the radial position of the cutoff layer, which typically increases as the density profile flattens. It has to be mentioned that uncertainties in the density profiles due to the dispersion in the Thomson scattering data are not systematically considered in this work. While their contribution may become relevant \cite{ref5}, it is not expected to alter significantly the presented results. In order to carry out a characterization of the dependence of the velocity shear magnitude with density and power under different configurations, we will define three characteristics of the profile as seen in the figure: in the first place, the variation of the radial electric field across the sign reversal, $\Delta$E$_r$, and the average value of E$_r$ in the SOL, E$_{r,SOL}$. Given the very short radial scale of the E$_r$ variation around the separatrix (which is substantially smaller than the separation between TS channels in that region), a precise calculation of the radial derivative of E$_r$ is considered unreliable. As well, the finite radial resolution of the DR can't resolve such narrow structures properly, thereby overestimating E$_r$ gradient values \cite{Happel10}. Therefore, $\Delta$E$_r$ will be used as a proxy for the velocity shear magnitude across this work. Secondly, as already discussed, the probing frequency at which the mentioned sign reversal is observed can be used along with a density profile from the TS in order to determine the radial position of the shear layer, $\rho_{shear}$.\\

Finally, in order to discuss the measured E$_r$ profiles, they are compared to the corresponding neoclassical calculation of E$_r$ carried out with the DKES \cite{ref4b} and KNOSOS \cite{ref4} codes for a number of relevant cases. For this, a database of DKES monoenergetic coefficients is convoluted with a Maxwellian distribution. At low collisionalities, DKES coefficients are replaced with KNOSOS calculations, in order to take properly into account the tangential magnetic drift \cite{ref4c}. For this, n$_e$, T$_e$ and T$_i$ profiles are taken as an input. The first two are obtained from the aforementioned Thomson scattering, while the later is obtained from the XICS diagnostic \cite{Novi14}. It must be taken into account that the experimental E$_r$ values correspond to the local value at the DR measurement position (roughly at the outer midplane of the elliptical section). For the purposes of this work, the electrostatic potential, $\varphi$, can considered to be constant on the flux-surface \cite{ref4d} and the local radial electric field to depend on the poloidal and toroidal position only through the flux expansion/compression: 

\begin{equation}
\left|E_r^{DR}\right|=-\left|\nabla \varphi\right|=-\frac{d\varphi}{dr}\left|\nabla r\right|,
\end{equation}

where $r\equiv a\rho$, being $a$ the minor radius. Therefore, the local radial electric field measured by the Doppler reflectometer, $E_r^{DR}$ must be normalized using $\nabla r$ in order to be compared to the neoclassical codes, which provide predictions of $d\varphi/dr$.\\

\section{Parametric study}\label{param}

\begin{figure}
	\centering
	\includegraphics[width=\linewidth]{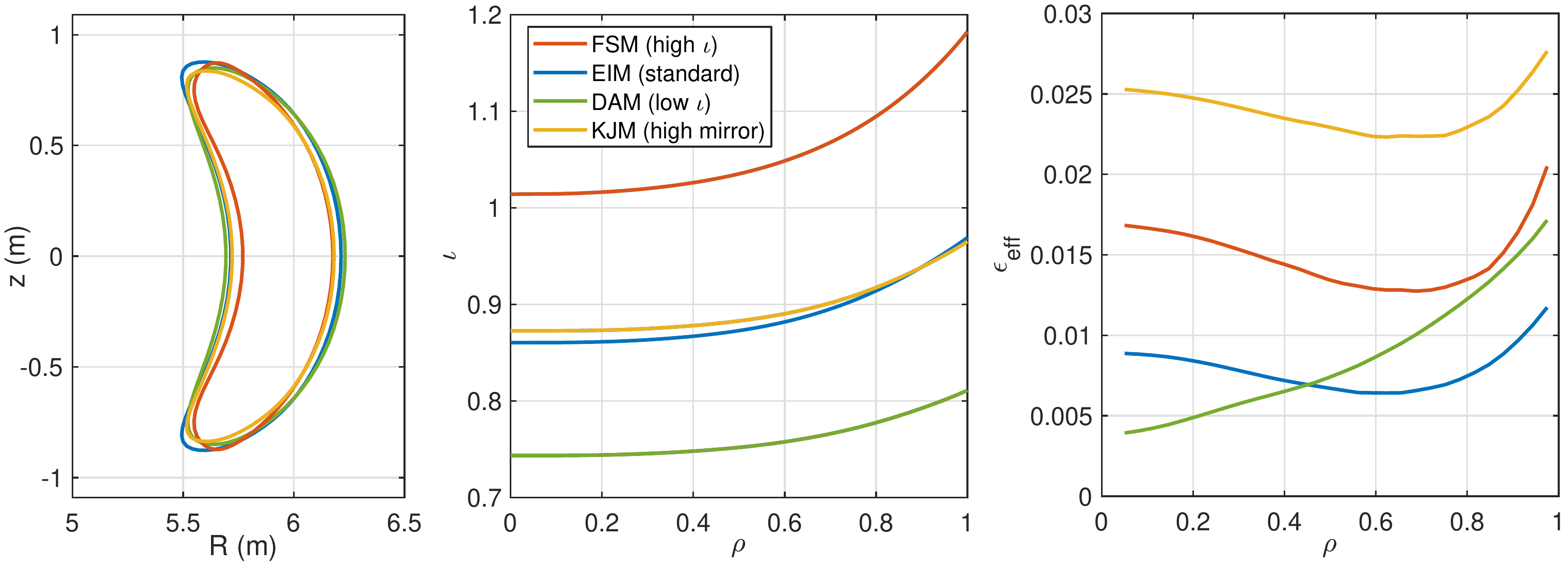}
	\caption{\textit{Overview of the magnetic configurations used in this work. Left: cross section of the LCFS at the DR toroidal position. Center: $\iota$ profile. Right: Effective ripple, $\epsilon_{eff}$, profile.}}
	\label{fig_conf}
\end{figure}

\begin{figure}
	\centering
	\includegraphics[width=0.5\linewidth]{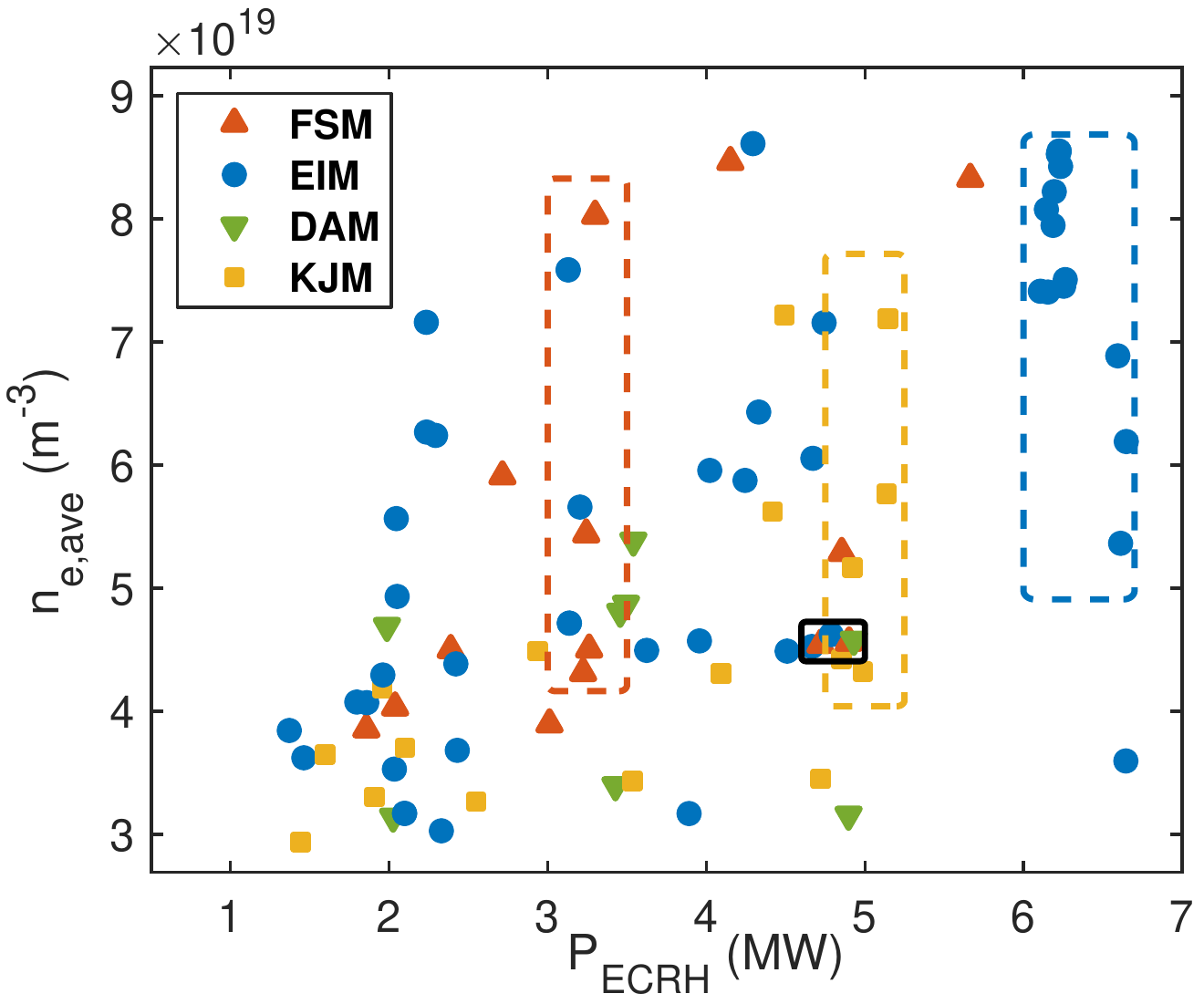}
	\caption{\textit{Parameter space. Blue/yellow/red/green symbols represent data samples in EIM/KJM/FSM/DAM configurations for different values of line average density, n$_{e,ave}$, and ECRH heating power, P$_{ECRH}$. Colored dashed boxes indicate samples used for the density scan displayed in figure \ref{fig2}. Black box indicates data points used in comparison to simulations in figure \ref{fig3}.}}
	\label{fig1}
\end{figure}

In order to study systematically the parametric dependencies of radial electric fields, we selected a number of representative radial profiles from DR measurements carried out in discharges with the most frequent magnetic configurations. These include the low $\iota$ configuration (DAM), standard configuration (EIM) and high $\iota$ configuration (FSM), featuring increasing $\iota$ profiles, decreasing bootstrap currents and relatively similar effective ripple, and the high mirror configuration (KJM), featuring an $\iota$ profile of intermediate value (very similar to the one of EIM), low bootstrap and substantially increased ripple. An overview of the different configurations used in this work is provided in Figure \ref{fig_conf}. More detailed information on the characteristics of the magnetic configurations used in this experimental campaign can be found in the literature \cite{ref3,Beidler11}. Discharges were chosen with the aim of covering the widest possible range of line averaged density values (as measured by interferometry \cite{Brunner18}) and ECRH power values for each configuration. The resulting parameter space is displayed in figure \ref{fig1}. As can be seen, the data obtained in the standard configuration (EIM, the most frequent during the campaign) covers most of the operational space in the campaign \cite{ref3c}. Instead, for the high mirror (KJM), high $\iota$ (FSM) and low $\iota$ (DAM) configurations, only partial coverage of such space results from the limited availability of data.\\

\begin{figure}
	\centering
	\includegraphics[width=0.6\linewidth]{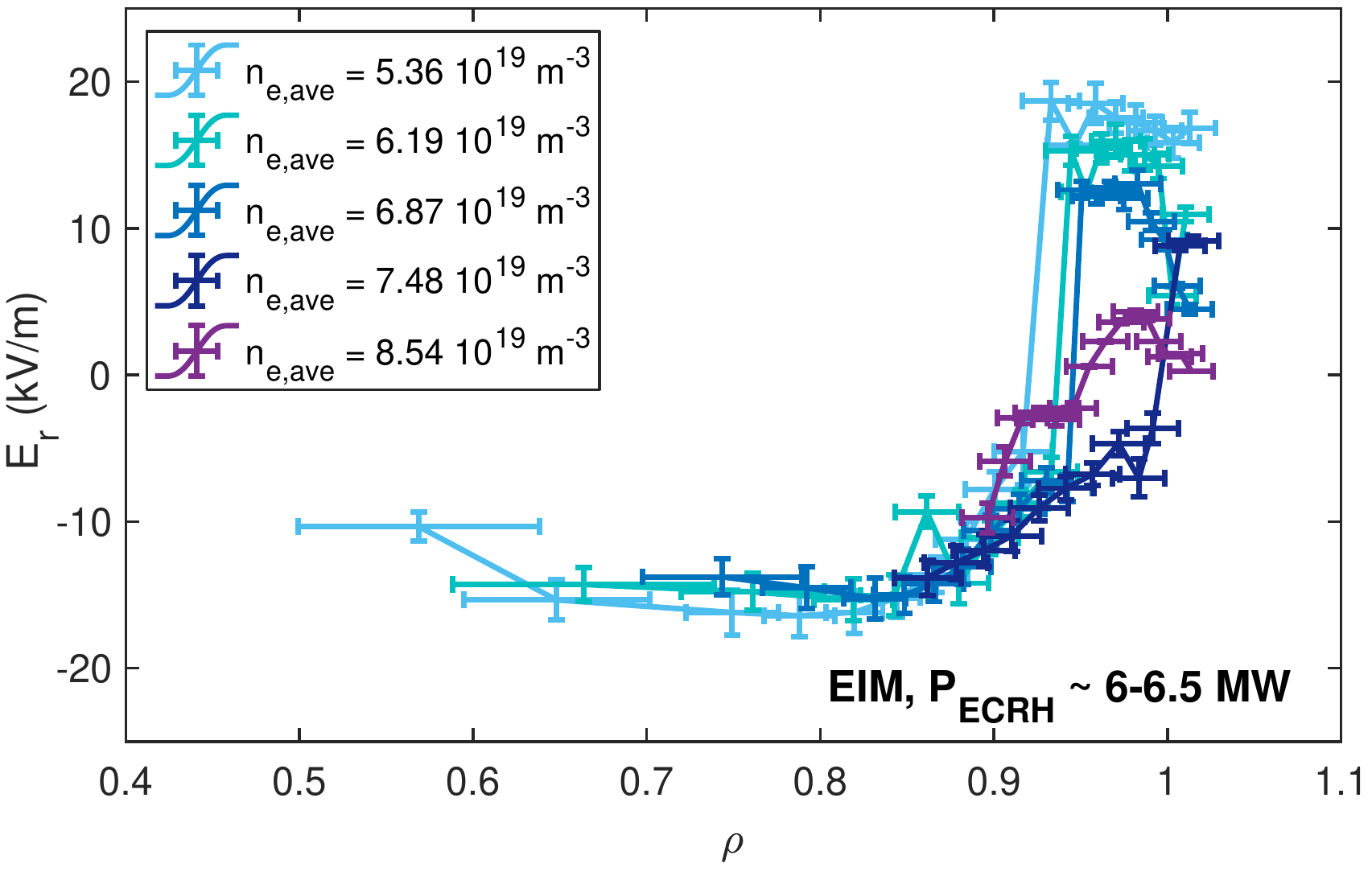}
	\includegraphics[width=0.6\linewidth]{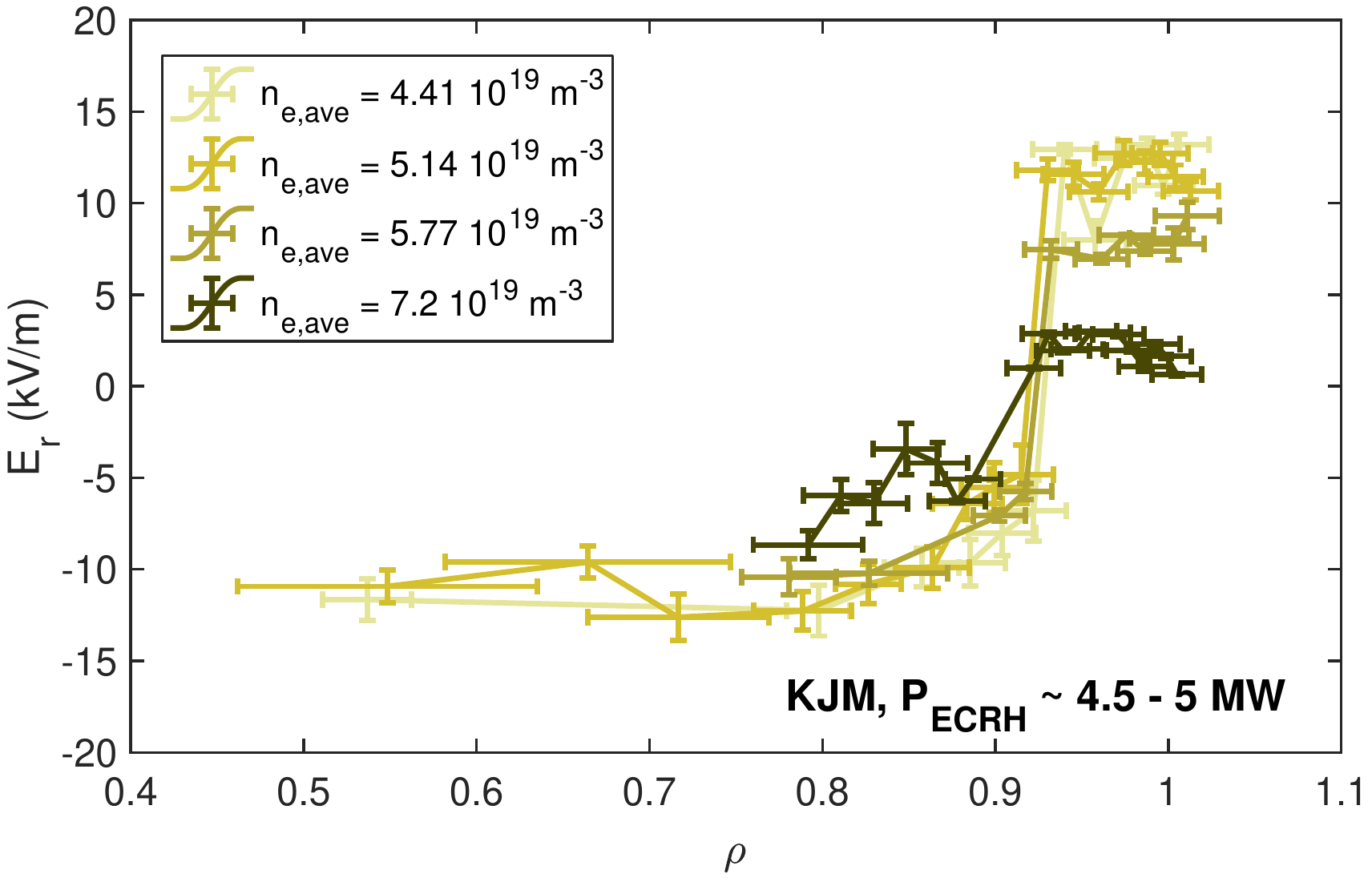}
	\includegraphics[width=0.6\linewidth]{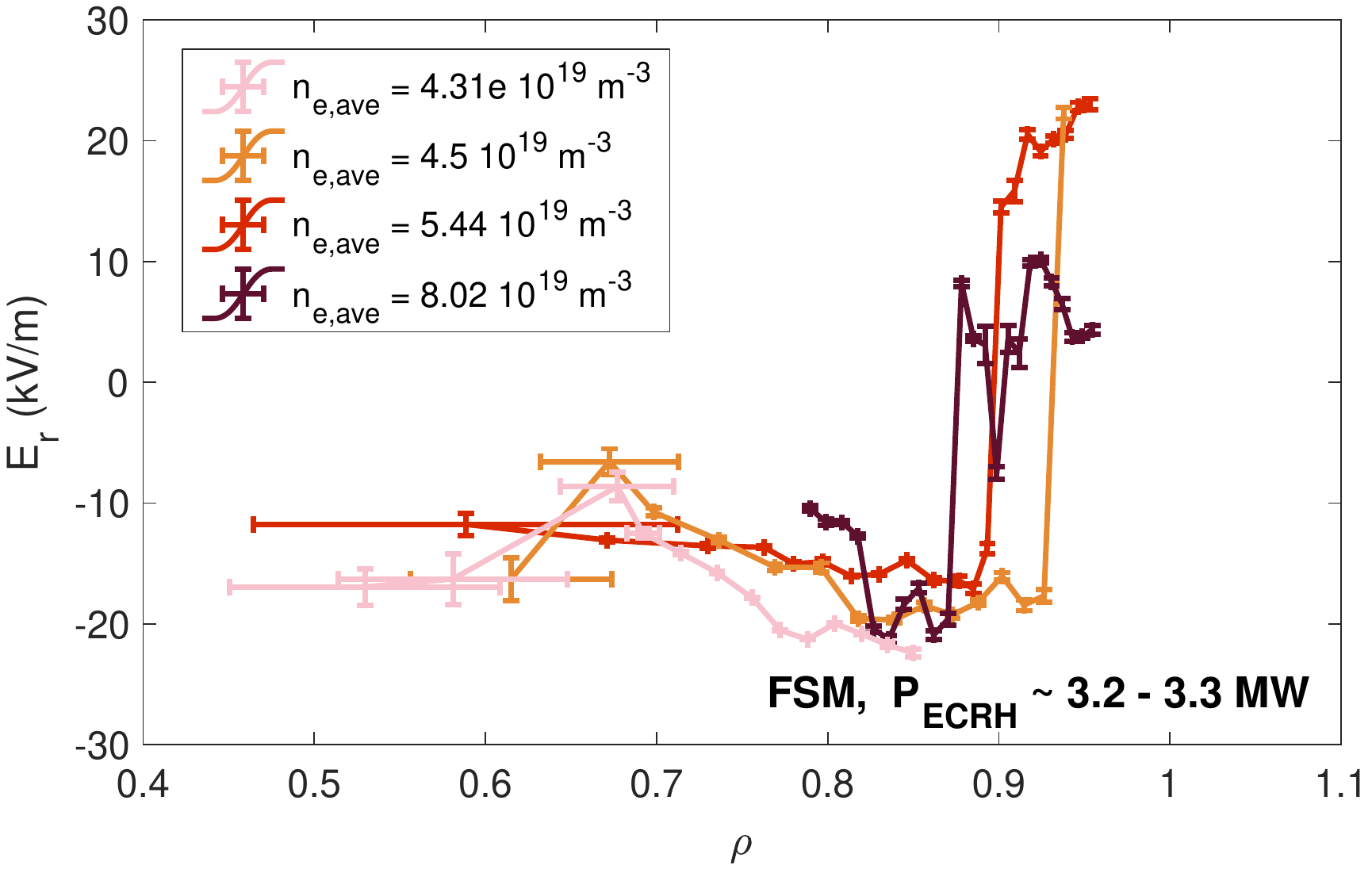}
	\caption{\textit{Density scan of the radial electric field profile. Top/middle/bottom display EIM/KJM/FSM configuration discharges.}}
	\label{fig2}
\end{figure}
 
\begin{figure}
	\centering
	\includegraphics[width=0.5\linewidth]{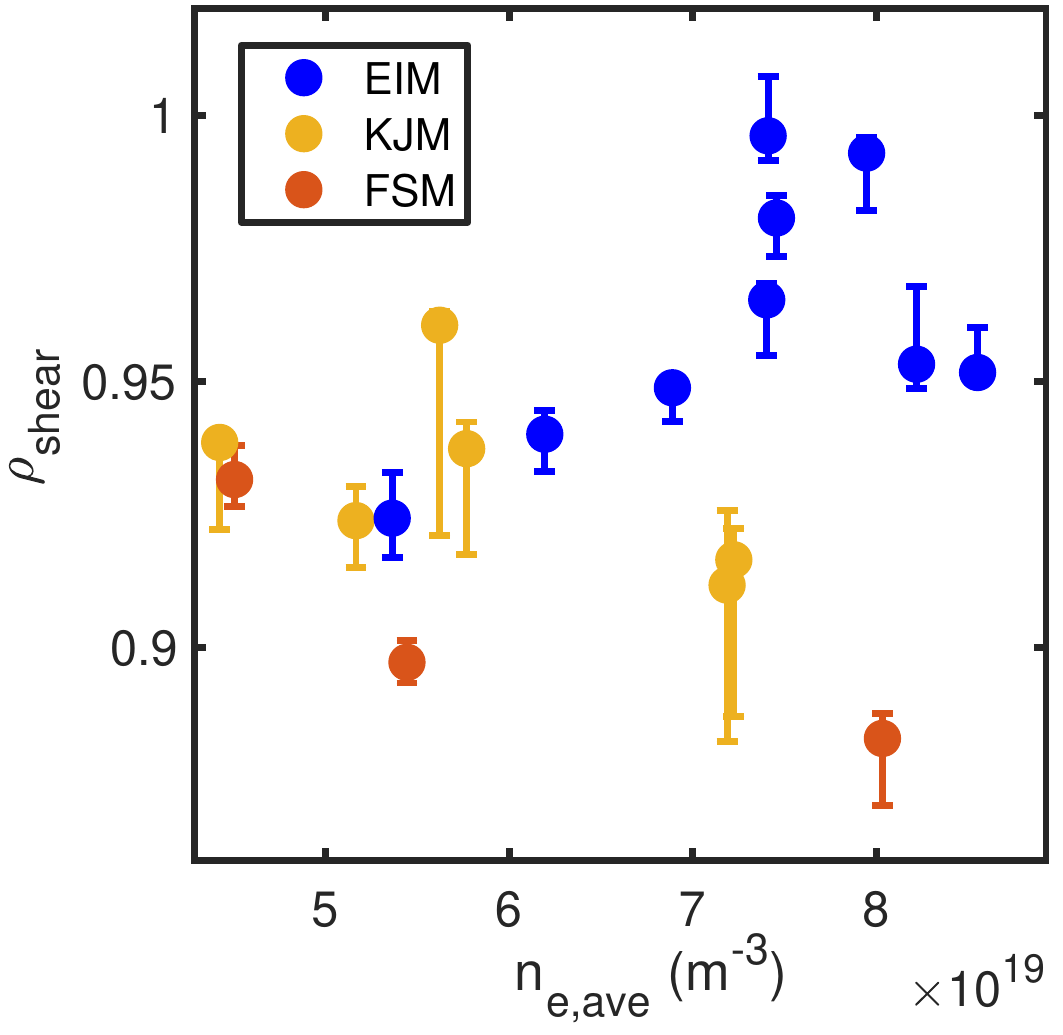}
	\caption{\textit{Position of the E$_r$ shear in the density scans at constant power displayed in figure \ref{fig1}.}}
	\label{fig2b}
\end{figure}

As a first step, a density scan is carried out by selecting a number of data points covering the whole range of density with a roughly constant heating power for EIM, KJM and FSM configurations. These are indicated by colored boxes in figure \ref{fig1}. Unfortunately, not enough data were available to run this scan for the same P$_{ECRH}$ values in all three configurations. In figure \ref{fig2}, the E$_r$ profiles corresponding to a selection of these points are displayed in order to show their dependency on density. As can be seen, the most prominent feature of the profiles is the previously discussed sudden reversal of E$_r$ sign taking place near the separatrix. Instead, E$_r$ in the confined region does not seem to change substantially with density (this contrast with some observations carried out in other machines \cite{Ida05}). Plots in figure \ref{fig2} are consistent with expectations, as the E$_r < 0$ observed in the core changes into E$_r > 0$ when approaching the LCFS. As can be seen in figure \ref{fig2b}, the precise radial position seems to depend on the magnetic configuration: in KJM and FSM, the position seems to move slowly inwards in the $\rho \simeq 0.9-0.95$ range as density increases. Instead, a different effect seems to be observed for the position of the sign reversal in EIM: starting from a similar $\rho \simeq 0.9$, the position increases with density until $\rho \simeq 1$ is reached for a line averaged value of n$_{e,ave} \simeq 7.5-8$ $10^{19} $m$^{-3}$. As will be discussed in Section \ref{SL}, this density corresponds roughly to the one in which detachment begins for this value of P$_{ECRH}$. For higher density values, the sign reversal position seems to move back towards the core.\\

\begin{figure}
	\centering
	\includegraphics[width=\linewidth]{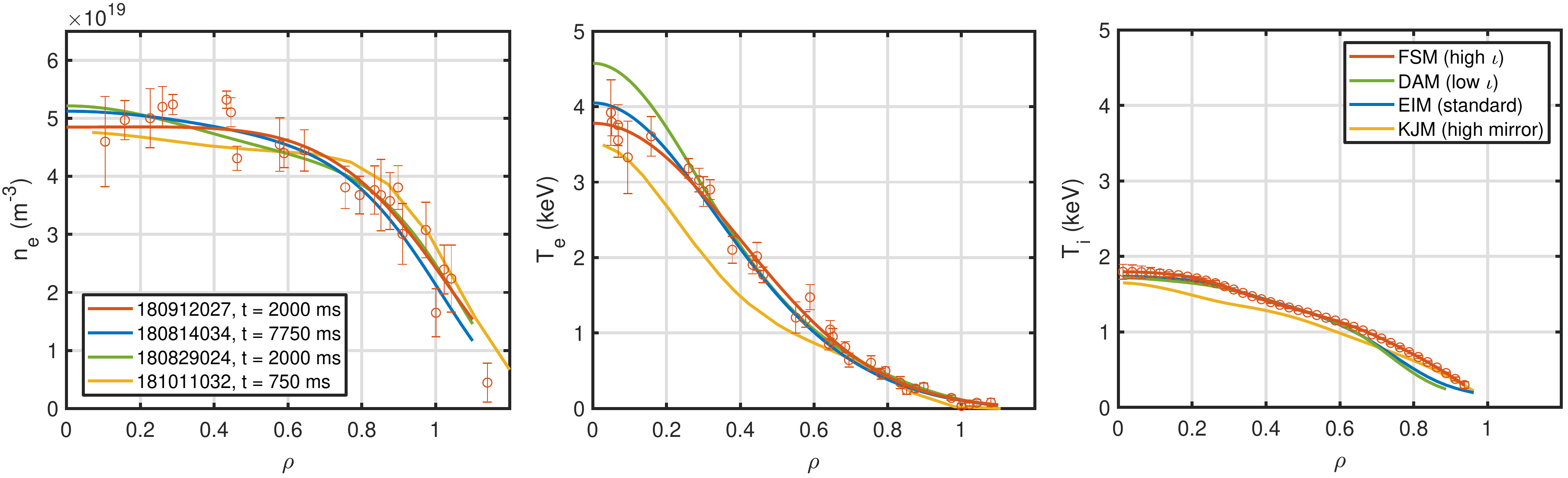}
	\caption{\textit{Fits for the density (left), electron temperature (center) and ion temperature (right) profiles for the four discharges contained in the black square in figure \ref{fig1}. Data points, as measured by the TS and XICS diagnostics, are included for shot 20180912027 as reference.  Colors indicate the magnetic configuration.}}
	\label{fig2e}
\end{figure}

\begin{figure}
	\centering
	\includegraphics[width=0.75\linewidth]{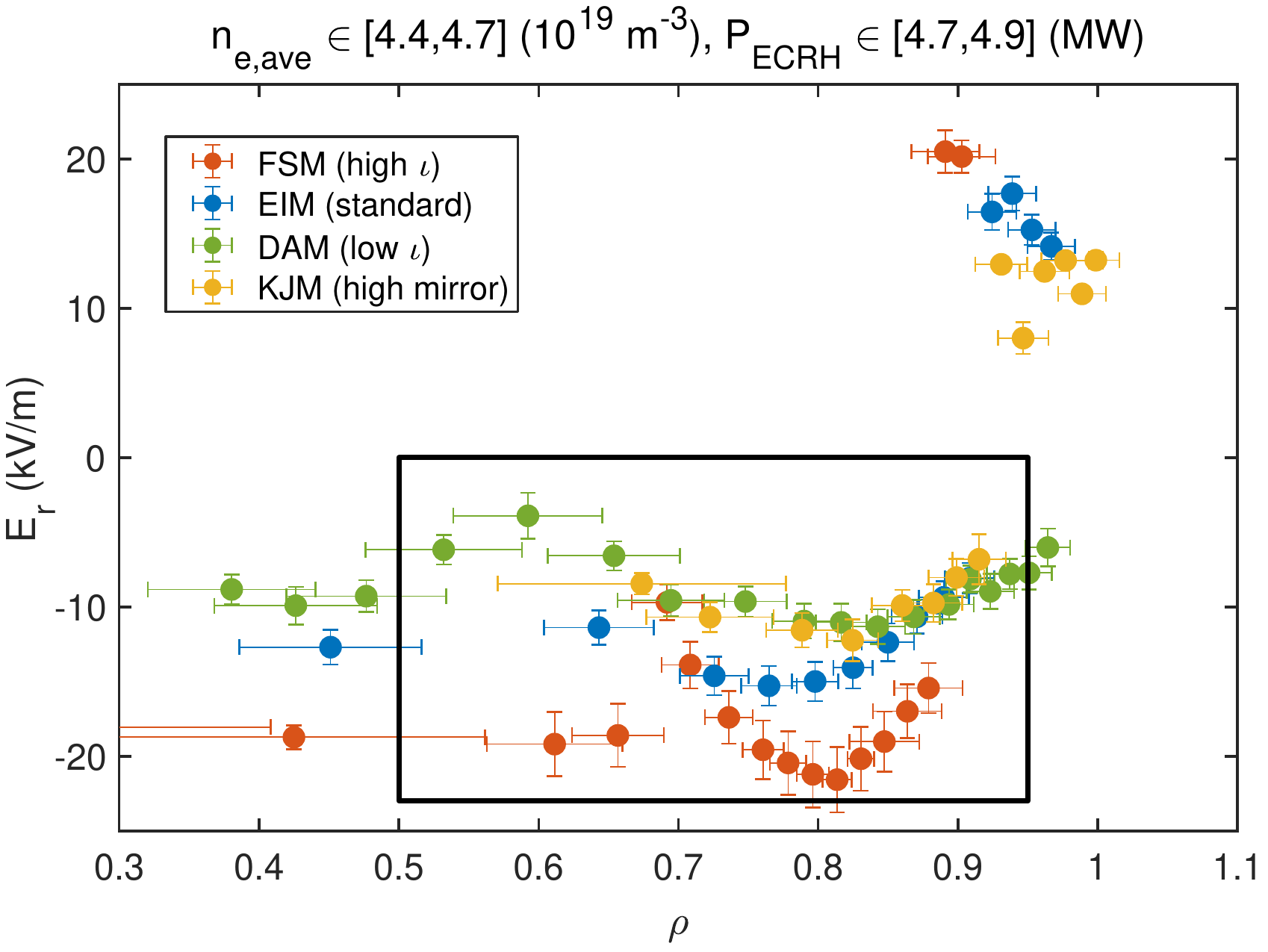}
	\includegraphics[width=\linewidth]{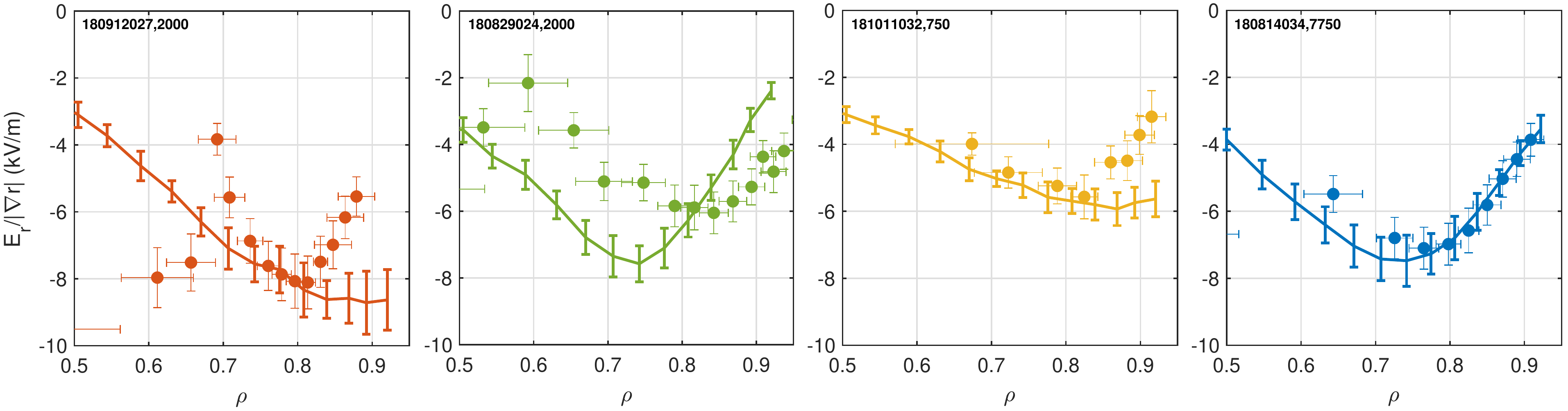}
	\caption{\textit{Top: average radial profile of E$_r$ for several magnetic configurations. Bottom: neoclassical simulations (thick lines) are compared to normalized E$_r$ profiles in the radial domain highlighted by the black square in the top figure. Colors indicate the magnetic configuration as in figure \ref{fig1} and \ref{fig2e}.}}
	\label{fig3}
\end{figure}

Next, a set of discharges with close values of density and heating power are selected to analyze the differences in the E$_r$ profile associated to the magnetic configuration. To this end, as shown by the small black box in figure \ref{fig1}, a parametric region is defined in which one sample of each configuration can be found within a narrow range of n$_{e,ave} \simeq 4.6 \cdot 10^{19} $m$^{-3}$ and  P$_{ECRH}\simeq 4.8 $ MW values. As can be seen in figure \ref{fig2e}, this similarity is essentially confirmed when looking at temperature and density profiles, which follow the same general trends. The fact that all four profiles are similar is consistent with the experimental observation that, in last campaigns, global confinement does not seem to depend substantially on the magnetic configuration \cite{Dinklage18,bozhenkovAPS}. Radial electric field profiles are displayed together in figure \ref{fig3}, where they share the same general characteristics of the profiles already shown in figure \ref{fig2} for the equivalent densities. Given the similarity between the profiles, differences between them can be attributed mainly to magnetic configuration effects: in particular, the edge E$_r$ value in the confined region seems to depend very clearly on the magnetic configuration. One trend that that can be observed is a dependency on the $\iota$ value of the configuration. Indeed, in all three cases, the modulus of the radial electric field increases with the rotational transform value: $|$E$_{r,DAM}|< |$E$_{r,EIM}|< |$E$_{r,FSM}|$. However, the KJM case (with an $\iota$ value similar to that of EIM) is not following this trend.  \\
%The only exception to this trend is the structure observed in the FSM configuration around $\rho = 0.7$ in which E$_r$ increases to values close to $-5$ kV/m. \\

\begin{figure}
	\centering		
	\includegraphics[width=0.5\linewidth]{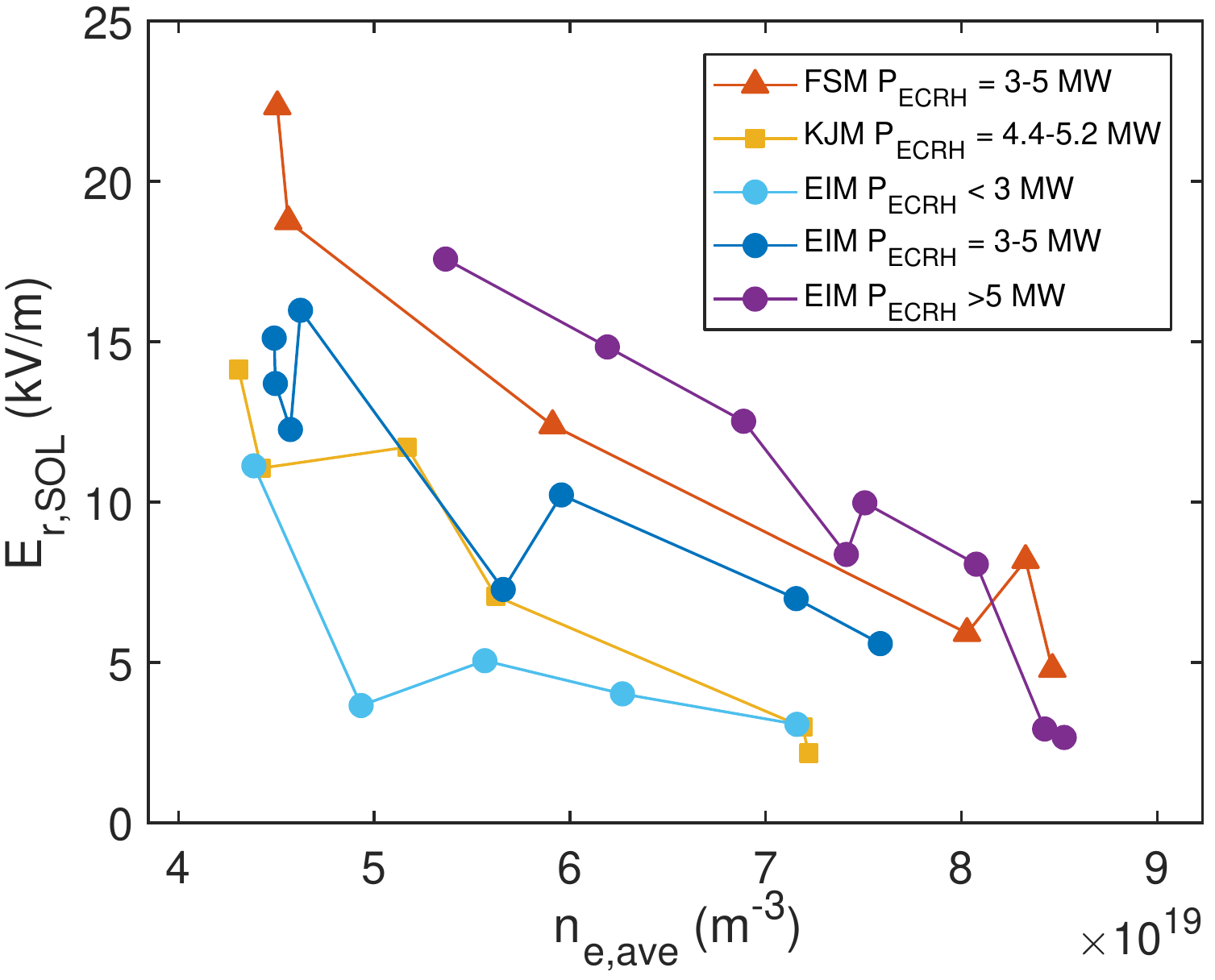}
    \includegraphics[width=0.5\linewidth]{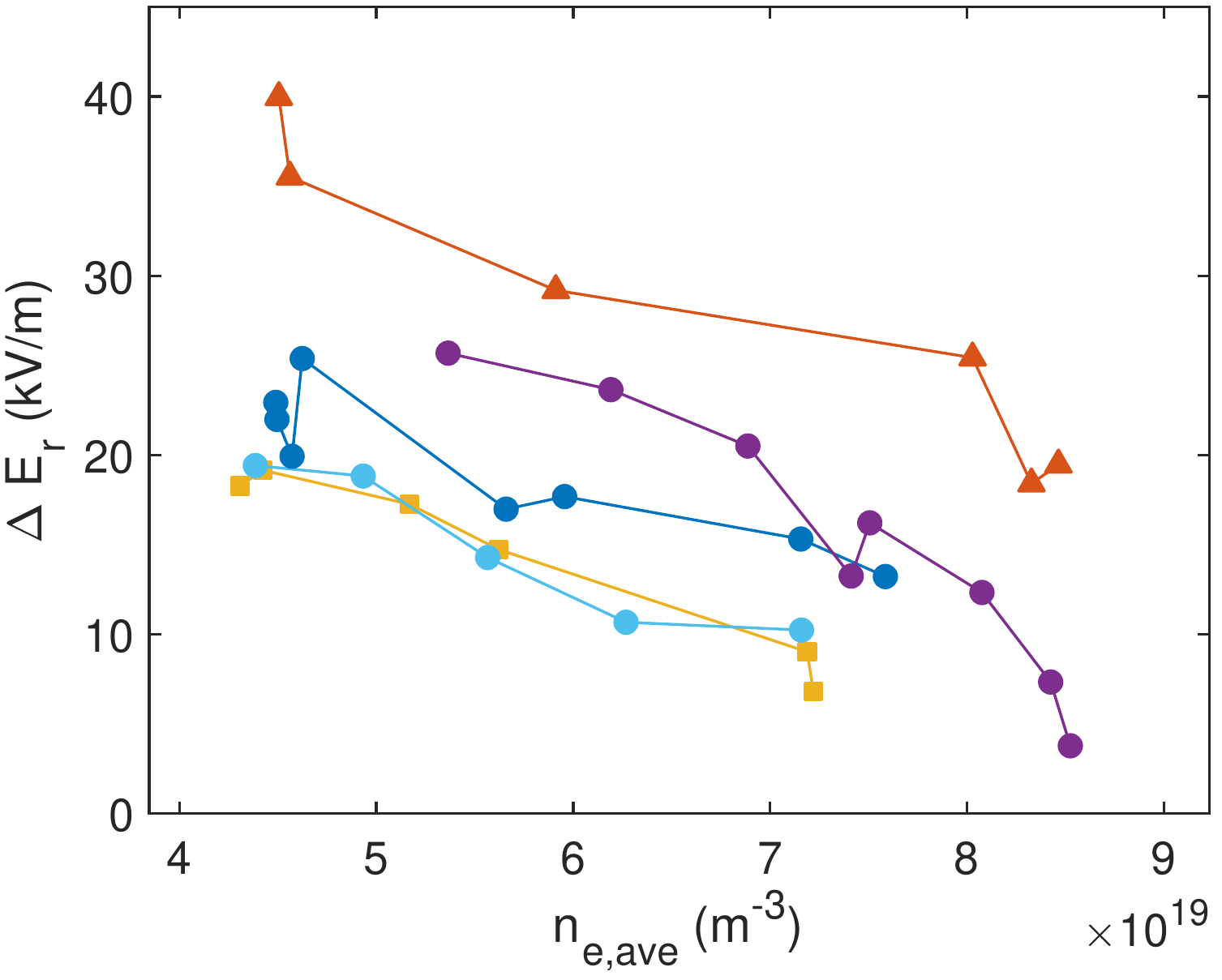}
	\caption{\textit{Dependence of E$_r$ at the SOL (top) and E$_r$ variation (bottom) with density for constant power. Red/blue/yellow colors correspond to high $\iota$/standard/high mirror configurations.}}
	\label{fig2c}
\end{figure}

Finally, the variation of $\Delta$E$_r$ and E$_{r,SOL}$ is evaluated over a larger sample of discharges, including different configurations, as well as different P$_{ECRH}$ values for the standard configuration EIM. Given that no discharge in the low $\iota$ configuration features a density sufficiently high for measuring fields in the SOL, that configuration is excluded from this analysis. The results are represented in figure \ref{fig2c}. In the top plot, a common feature to all configurations can be readily seen: E$_{r,SOL}$ decreases with density in all cases. In the EIM case, in which data featuring different values of P$_{ECRH}$ is available, E$_{r,SOL}$ increases with heating power. Both observations are consistent with the already discussed interpretation for the SOL positive E$_r$, as the peak value of T$_e$ at the target is expected to drop when density is increased with fixed heating, thereby reducing $\nabla$T$_e$ at the wall in the direction perpendicular to the strike line. Conversely, T$_e$ at the target is expected to increase when P$_{ECRH}$ is increased for constant densities. Not surprisingly, the behavior of the E$_r$ shear is then a combination of these tendencies and the previously discussed dependence of |E$_r$| with $\iota$ at the plasma edge: $\Delta$E$_r$ displays a very similar dependency with power and density as E$_{r,SOL}$. However, the effect of configuration is stronger, with high $\iota$ discharges displaying substantially larger shear than those of EIM and KJM as a result of the more negative electric field value in the confined region. Interestingly, these results are consistent with measurements at lower densities by correlation reflectometry \cite{AKF2019}, which reported for FSM discharges with n$_{e,int} \simeq 4 \cdot 10^{19} $m$^{-2}$ and P$_{ECRH} = 5$ MW, values of E$_{r,SOL} \simeq 30$ kV/m and $\Delta$E$_r \simeq 50$ kV/m. These substantially larger values of electric field seem to follow roughly the tendency with density observed in Figure \ref{fig2c}. While magnetic configuration seems to have a strong and general effect on the E$_r$ shear, there are at least some situations in which global confinement does not seem to depend strongly on it: as an example, the very similar profiles displayed in Figure \ref{fig2e} correspond to quite different values of  $\Delta$E$_r$ in Figure \ref{fig2c}. This preliminary result suggests that global confinement is not affected by the possible suppression of turbulence caused by  $\Delta$E$_r$ at the edge. However, a more systematic analysis of this issue is required in order to reach general conclusions. This is out of the scope of the present study and will be addressed in future work. \\

\section{Comparison to neoclassical simulations}\label{neo}

In order gain a better understanding of the E$_r$ measurements presented in the previous section, a selection of them are compared to simulations carried out using the neoclassical codes DKES and KNOSOS, which should be able to reproduce their parametric dependencies. In these runs, impurity profiles have not been taken into account. This should have little impact, as previous sensibility studies showed small dependency on $Z_{eff}$ \cite{AKF2019,Windisch17}. In order to account for the uncertainty of the measurements, a sensibility study has been run in each case assuming deviations of up to 10\% from the available density and temperature profiles, as well as their gradients. The standard deviation of the different resulting curves is represented as the error bars.\\

In the first place, the shots included in the configuration comparison (black box in figure \ref{fig1}) are considered and the neoclassical E$_r$ is calculated from their profiles, shown in figure \ref{fig2e}. The results are displayed in the bottom plots of figure \ref{fig3}, in which the experimental profiles of E$_r$ (normalized using $\left|\nabla r\right|$, as explained in the introduction)  are represented for the limited radial range in which the input profiles for neoclassical simulations are considered reliable ($\rho < 0.9$) while the density profile is steep enough to allow for a good radial localization in the DR measurements ($\rho > 0.5$) . As can be seen, flux expansion normalization partially reduces differences between high $\iota$ and standard configurations: the strongly negative E$_r$ values in FSM are partly due to local flux compression at cut-off position. Still, some of the $\iota$ ordering $|$E$_{r,DAM}|< |$E$_{r,EIM}|< |$E$_{r,FSM}|$ seems to persist after the normalization. On the other hand, the high mirror configuration (which features an $\iota$ profile similar the standard configuration) still shows the weakest electric field after the normalization, $|$E$_{r,KJM}|< |$E$_{r,DAM}|$, both in experiment and simulation. This indicates that the $\iota$ value is not in itself the most relevant parameter determining the E$_r$ difference between configurations, and that other effects (such as the different effective ripple) must be playing a role in it. Regarding the comparison between experiment and simulation, good agreement is found in three of the four configurations (EIM, FSM and KJM), including a quite accurate calculation of the minimum value of E$_r$. The agreement in the low $\iota$ configuration (DAM) is clearly worse, although the general trend of the experiment is recovered. \\
%As well, the aforementioned structure in the FSM configuration, is neither reproduced by the simulation and will require further analysis.\\ 

\begin{figure}
	\centering
	\includegraphics[width=0.75\linewidth]{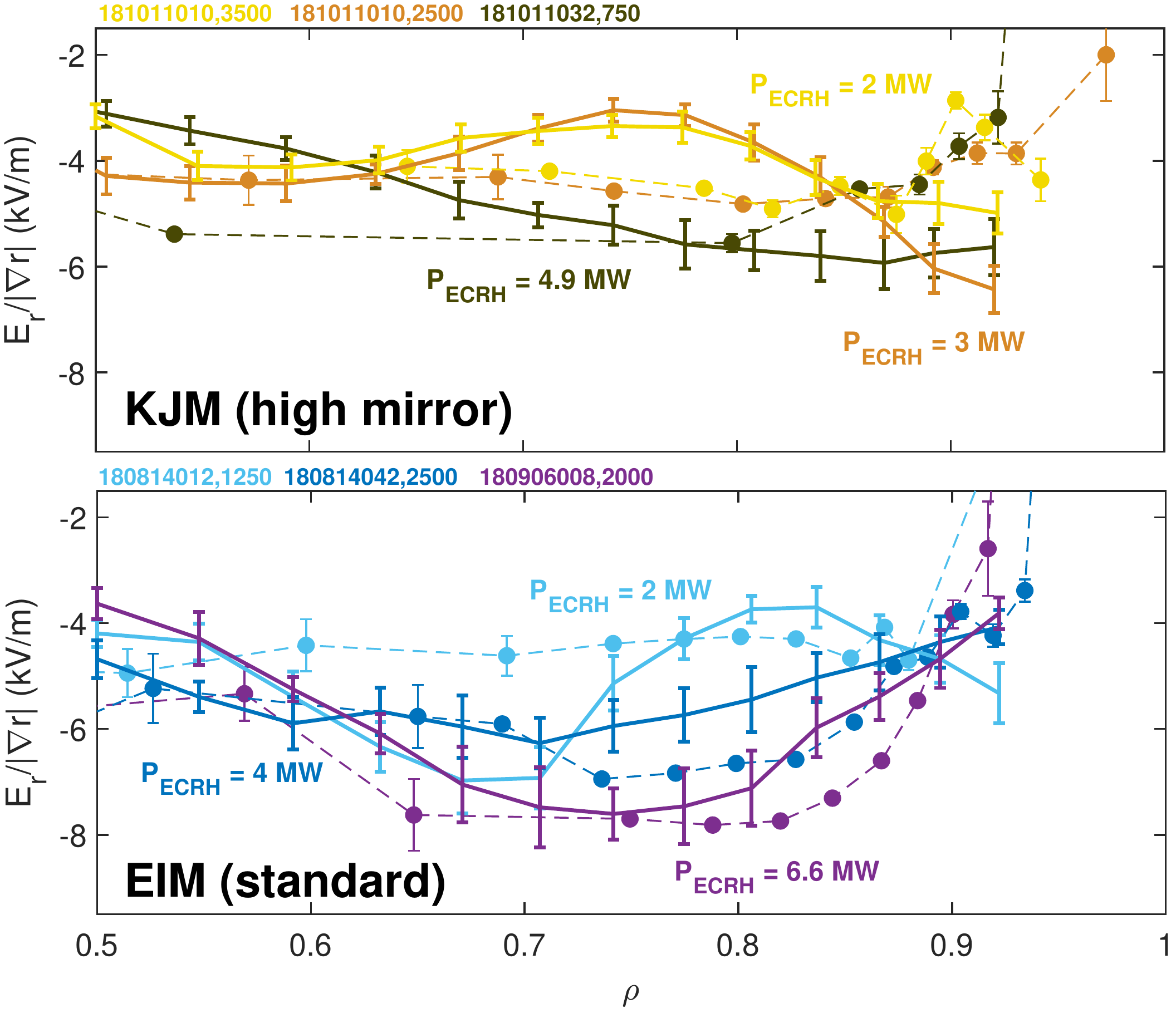}
	\caption{\textit{Comparison of P$_{ECRH}$ dependence of E$_r$ for KJM and EIM configurations. Dashed lines and circles represent DR measurements. Thick solid lines represent neoclassical simulations.}}
	\label{fig4}
\end{figure}

From this starting point, a parametric dependence study of the E$_r$ is carried out in this experiment-code overlap zone by analyzing the impact of changing the value of P$_{ECRH}$ in a number of discharges with similar values of n$_{e,ave}$. The most prominent result, displayed in figure \ref{fig4}, is that a clear dependence of E$_r$ on P$_{ECRH}$ in the $\rho \simeq 0.65-0.85$ region is observed in the experiment for the standard configuration: as can be seen, E$_r$ profiles in both configurations are similar for P$_{ECRH} \simeq 2$ MW (note that flux expansion does not play a role in this, since the profiles are normalized). However, as P$_{ECRH}$ is increased, the normalized value of the field almost doubles as P$_{ECRH}$ goes from $2$ to $6.6$ MW for the standard case. Instead, a clearly weaker tendency is found for the high mirror configuration, with no variation over a $1$ MW increase and a $25$\% change when the power is increased from $2$ to almost $5$ MW. This feature is qualitatively recovered by neoclassical simulations: in the standard case, simulations reproduce rather accurately not just the power dependency in EIM, but also the radial electric field values measured for each discharge. In the high mirror case, there is only a qualitative agreement with the data, but the variation of the predicted radial electric field with P$_{ECRH}$ is also clearly reduced with respect to the standard case. The different E$_r$ dependence on power of the configurations stems from two complementary sources: first, differences in neoclassical transport, which may lead to different values of E$_r$ for given plasma profiles; second, differences in the anomalous transport, which produces different plasma profiles for a given input power. The disentanglement of these two effects is left for future work, as it would require a more systematic comparison between discharges of the two configurations, including the local gradients of density and T$_i$ profiles and also to exclude other potential differences between them, such as P$_{rad}$, Z$_{eff}$, toroidal current, etc. 

\section{Evolution of E$_r$ during detachment}\label{SL}

\begin{figure}
	\centering
	\includegraphics[width=0.75\linewidth]{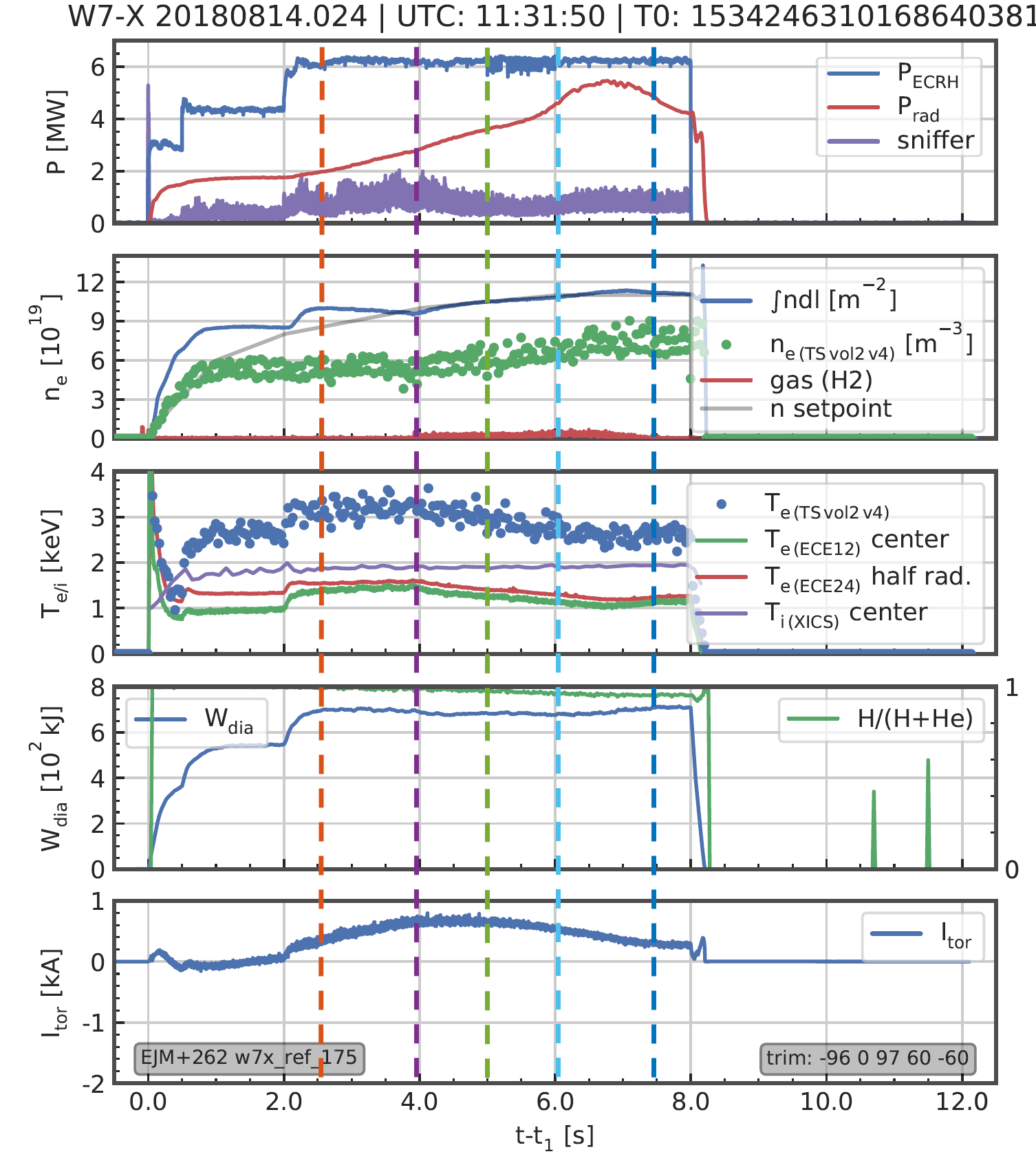}
	\caption{\textit{A typical standard configuration detached discharge. The five rows represent (from top to bottom) the evolution of power and radiation, line integrated density ($\int ndl = $L$_{inter}$n$_{e,ave}$, where L$_{inter} = 1.33$ m for the standard configuration), ion/electron temperatures, diamagnetic energy, $W_{dia}$, and toroidal current, I$_{tor}$. Vertical dashed lines indicate times selected for analysis.}}
	\label{fig5}
\end{figure}

\begin{figure}
	\centering
	\includegraphics[width=0.7\linewidth]{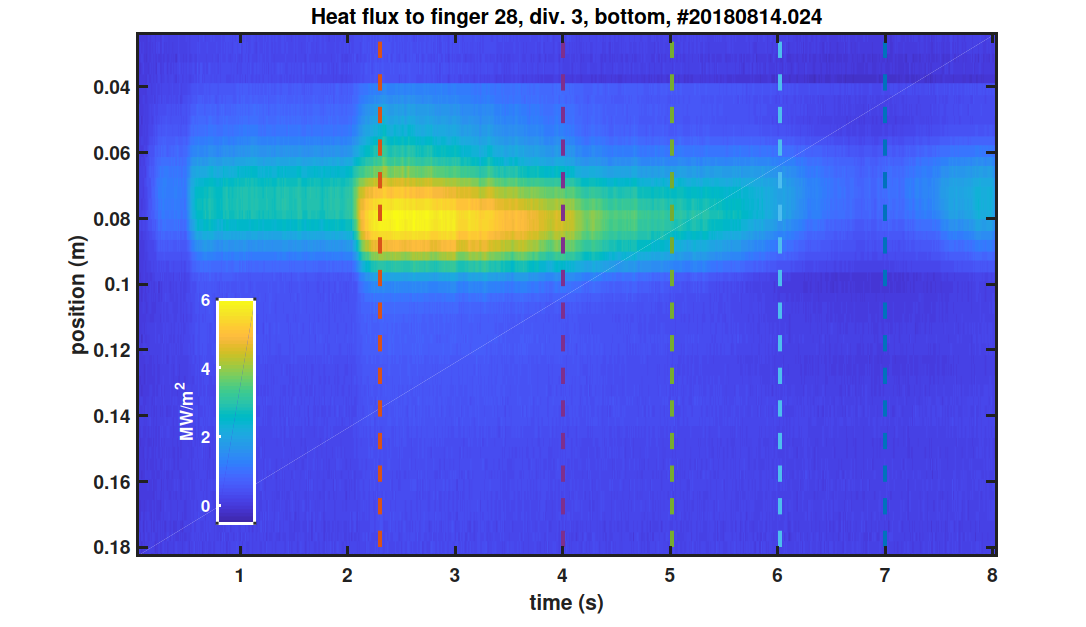}
	\caption{\textit{Infrared imaging of the divertor in shot 20180814024. The evolution of the heat flux in onto one of the targets is calculated from infrared camera data. Vertical axis shows the distance to the pumping gap across the target, perpendicular to the main direction of the strike line. The analysis times are marked with the same colors as in figure \ref{fig5}. }}
	\label{fig5b}
\end{figure}

\begin{figure}
	\centering
	\includegraphics[width=0.5\linewidth]{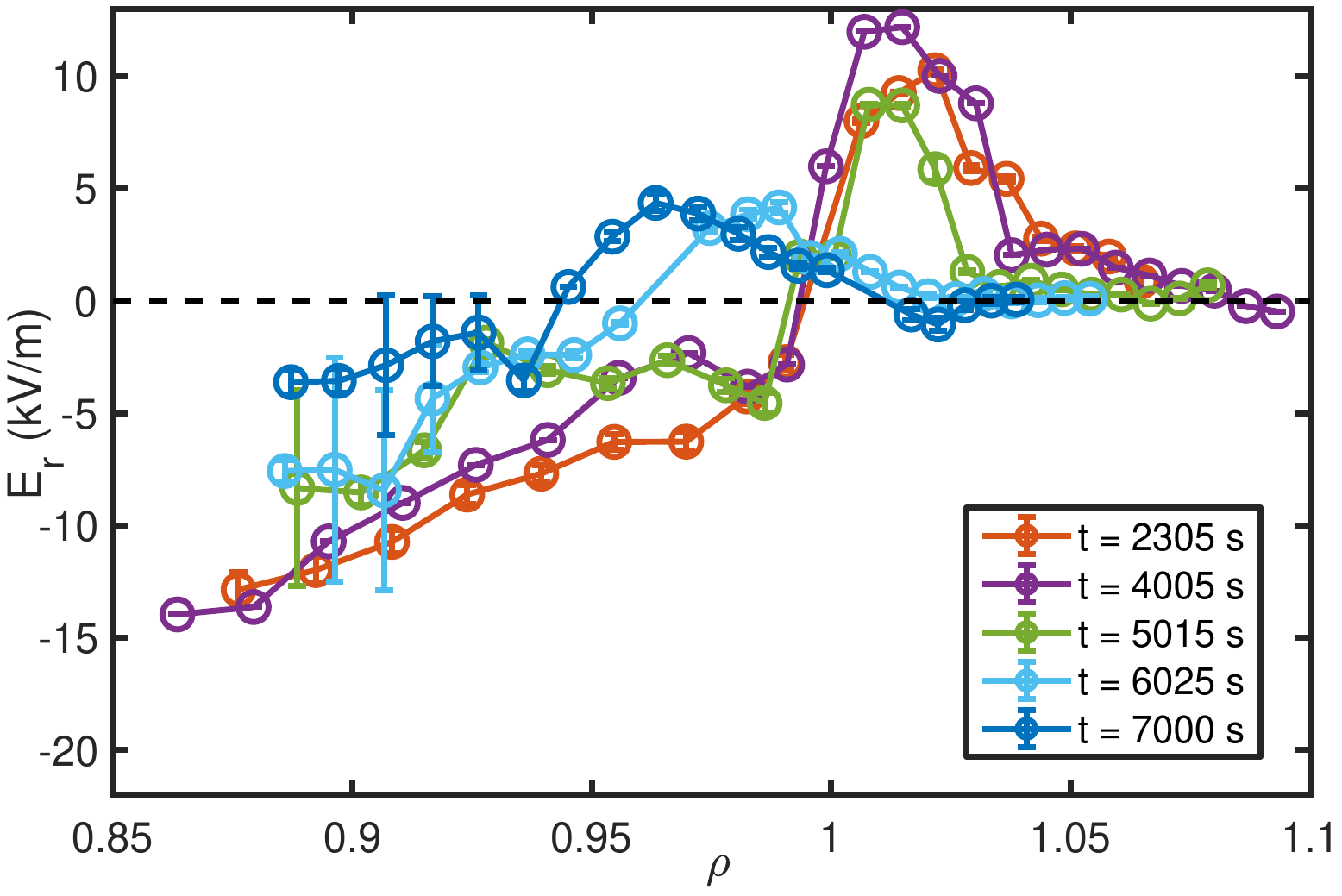}
	\caption{\textit{E$_r$ profiles in shot 20180814024. Colors correspond to the times indicated by dashed vertical lines in figure \ref{fig5}.}}
	\label{fig6}
\end{figure}

\begin{figure}
	\centering
	\includegraphics[width=0.9\linewidth]{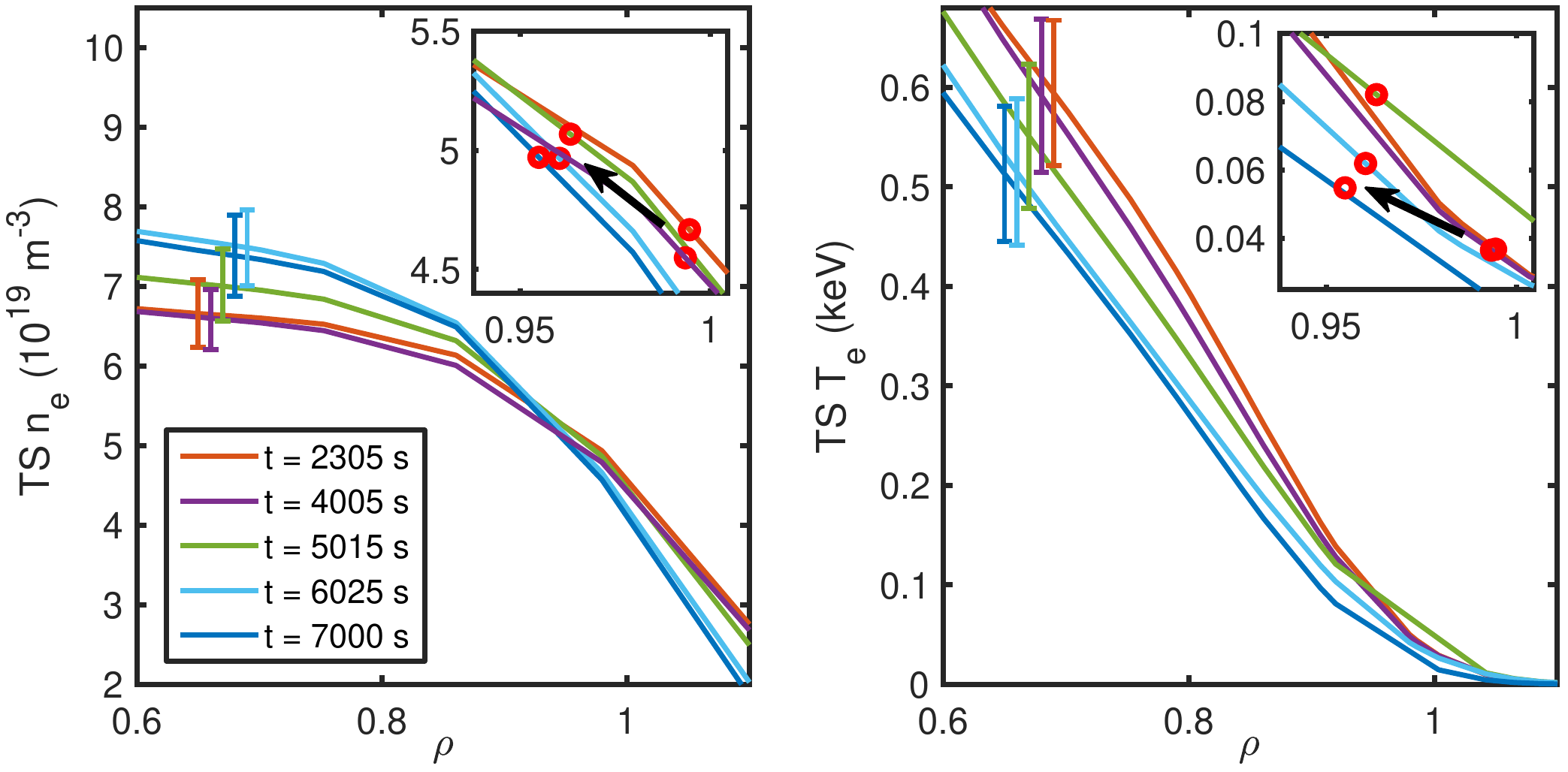}
	\caption{\textit{Density and electron temperature profiles in shot 20180814024. Colors correspond to the times indicated by dashed vertical lines in figure \ref{fig5}. Representative errorbars have been provided for each curve In the inserts, the position of the shear is shown  as a red circle in each curve. Black arrows indicate their evolution over time.}}
	\label{fig7}
\end{figure}

As mentioned in the introduction, one of the main milestones achieved during the 2018 campaign was the full divertor detachment at high density. A typical detachment process in standard configuration is displayed in figure \ref{fig5}: first, there is an attached phase for t $<2$ s, in which density is moderate and radiation represents a small fraction of the injected power. Then, after P$_{ECRH} = 6$ MW is achieved at t $=2$ s, density is increased leading to a gradual decrease of $T_e$ and eventually to the detachment when $n_{e,line} \simeq 11 \cdot 10^{19} $m$^{-2}$ (corresponding to an averaged value of $n_{e,ave} \simeq 8.25 \cdot 10^{19} $m$^{-3}$) and P$_{rad}$ approaches the value of the injected power. The onset of detachment can be followed in figure \ref{fig5b}, in which the heat flux onto one of the divertor targets is displayed, as calculated from infrared imaging. In it, it can be seen how the heat flux mirrors the trend in P$_{rad}$ starting to decrease for t $ > 2.5$ s and reaching a minimum around t $ = 7$ s, in which power detachment can be considered almost complete. In order to evaluate the evolution of E$_r$ over the detachment, five times (indicated in the figures as vertical dashed lines) are selected at different stages of the process.  The results are displayed in figure  \ref{fig6}: as the density is increased, and in agreement with the results discussed in section \ref{param}, E$_r$ in the SOL drops with increasing density. This is consistent with the reduction in T$_e$ observed at the targets. As well, following the trend already displayed in figure \ref{fig2b} for high densities, the density at the shear position increases over that measured by the TS at the separatrix, indicating that the former is moving radially inwards. This can be seen more clearly in figure \ref{fig7}, in which the average TS profiles are displayed for the DR ramps corresponding to the five times indicated in figure \ref{fig5}: during the density ramp leading to detachment, TS data shows an increase of n$_e$ inside the LCFS, which results in a stepper profile. Since the stored energy in the plasma remains roughly constant, a general drop in edge T$_e$ compensates that increase of density. As explained in the Introduction, this implicates that global confinement is not affected by the increase of P$_{rad}$ \cite{Jakub20}. The reason for this is probably that most of the radiation remains in the SOL, even for high values of P$_{rad}$/P$_{ECRH}$ \cite{Feng16,Pedersen19}. This process contrasts with the RMP-induced detachment achieved in LHD, in which the electron pressure profile becomes more peaked and W$_{dia}$ and $\tau_E$ increase during the density ramp (albeit substantially reduced by the global volume contraction caused by the RMP) \cite{Kobayashi19}. Meanwhile, the position of the shear (indicated in the insert as red circles on top of each profile) moves radially inwards, into regions of higher n$_e$ (the evolution of T$_e$ remains clearly within the errorbars. However, in figure \ref{fig9} it will be shown how T$_e$ follows a similar trend).\\

In order to carry out a more systematic analysis, we have expanded this manual evaluation of a reference discharge using an algorithm \cite{ref9} to detect automatically the shear region local density from the DR data: as discussed in the introduction, the density at the shear layer can be calculated from the DR probing frequency for which the E$_r$ sign reversal is observed. Although it is not always possible to carry out a precise detection of the Doppler shift automatically, relatively simple techniques (such as the calculation of the center of gravity of the spectrum) suffice for this analysis, as only the time for which the Doppler shift changes sign needs to be identified. This automatic analysis has been applied to nearly $20$ standard configuration discharges taken from a session dedicated to stabilized detachment (comprised in the 181010007-040 interval), mostly featuring density ramps at high power (P$_{ECRH} \simeq 5-6$ MW), similar to the previously discussed discharge 180814024. As well, one low power-detached discharge has been included, in which an equivalent level of P$_{rad}$ has been achieved by means of Nitrogen seeding (180920050). Finally, another two discharges (181010031, 181010037) have been considered in which power is reduced after detachment and density is brought down by feedback control in order to keep detached conditions. An example of this is displayed in figure \ref{fig5c}, in which the evolution of power and corresponding reduction of density can be seen starting at t $\simeq 4.5$ s. By this increase in the sample size, the importance of several error sources (most importantly, the dispersion on density profiles, seen in figure \ref{fig2e}) is reduced, revealing underlying statistical trends. In this automated analysis, the shear position has been evaluated with respect to two parameters driving detachment: average density and radiated fraction of power, P$_{rad}$/P$_{ECRH}$. As well, the evolution of the plasma shape is estimated introducing the form factor $F$=n$_{e,ave}$/n$_{e,shear}$, indicating the ratio between line averaged density and density at the shear. Assuming that n$_{e,shear} \simeq $n$_{e,LCFS}$, an increase of $F$ indicates a transition towards a more peaked plasma profile. \\

\begin{figure}
	\centering
	\includegraphics[width=\linewidth]{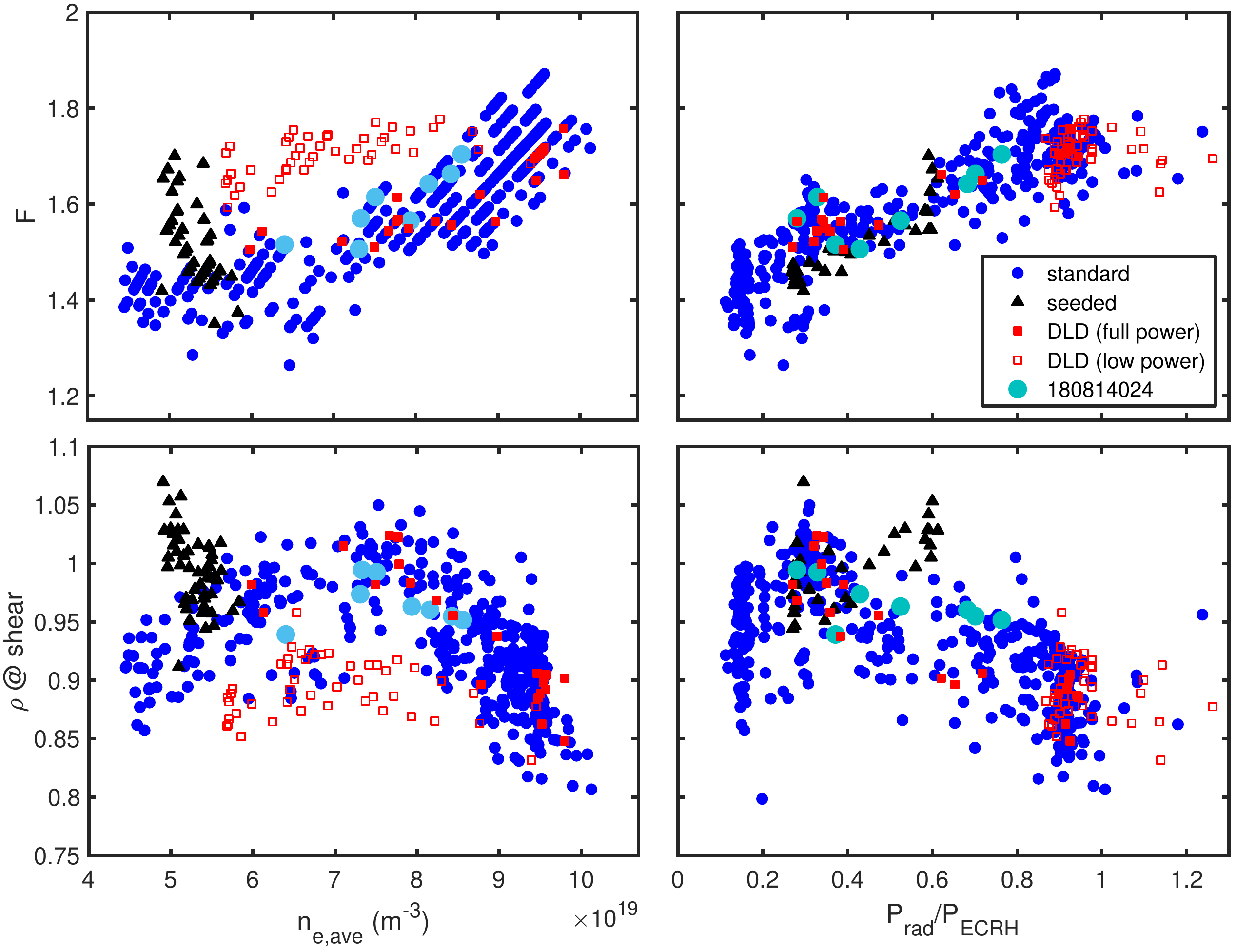}
	\caption{\textit{ Results of the automatic analysis. On the top and bottom row shape factor $F=$n$_{e,ave}$/n$_{e,shear}$, and position of the shear, respectively. Conversely, left and right columns are represented as a function of averaged density and radiated fraction of power. Colors and symbols indicate the group of discharges as explained in the text, including standard density ramp scenario, seeded detachment, detachment at low density (DLD, with solid/hollow points indicating times before/after the power reduction, as seen in the example in Figure \ref{fig5c}) and time points from Figure \ref{fig5}, corresponding to shot 180814024.}}
	\label{fig8}
\end{figure}

\begin{figure}
	\centering
	\includegraphics[width=0.75\linewidth]{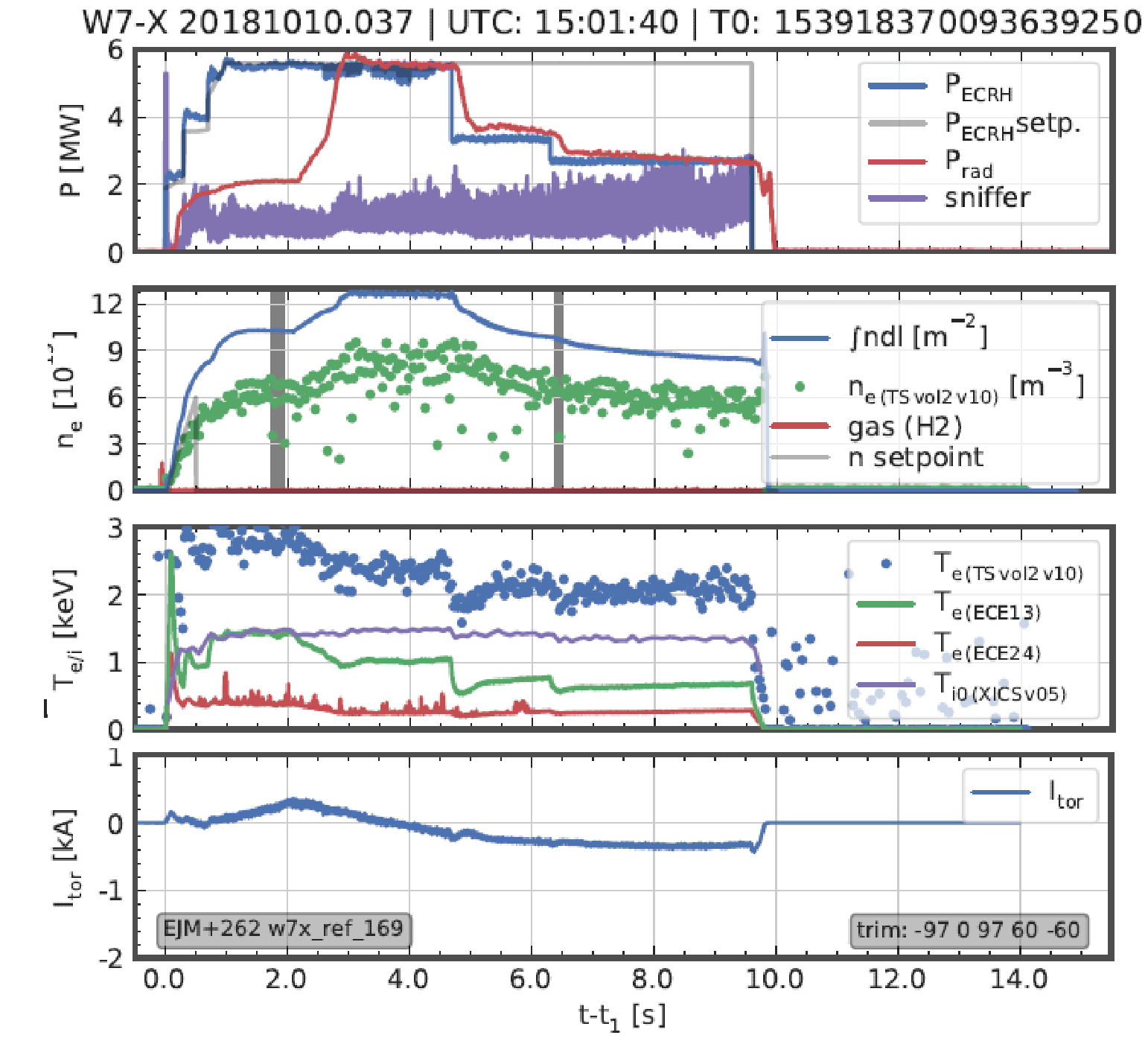}
	\caption{\textit{Main traces of discharge 20181010037, featuring a controlled reduction of P$_{ECRH}$ after detachment is achieved, keeping P$_{rad}$/P$_{ECRH}\simeq$ 1. As in figure \ref{fig5}, the four rows represent (from top to bottom) the evolution of power and radiation, density, ion/electron temperatures and toroidal current.}}
	\label{fig5c}
\end{figure}

The main results of the analysis are presented in figure \ref{fig8}, with symbols and colors indicating each of the discussed groups of discharges: base density ramp scenario (blue circles), seeded detachment (black triangles) and power reduction after detachment (red squares). In the last case, solid/hollow symbols were used to indicate times before/after the power reduction. Finally, as a test for the automatic analysis, values obtained manually in reference discharge 180814024 are also included as large, light blue circles. As can be seen in the top left plot, $F$ remains roughly constant for low density values, and increases with density for n$_{e,ave} > 7.5 \cdot 10^{19} $m$^{-3}$, corresponding to the onset of detachment. A similar trend is followed by solid red points, but then a higher value of $F$ is conserved when power is reduced while retaining detachment, indicating that the change in shaping is related to the detached condition rather than to density. In the seeded case, the high $F$ values corresponding to detachment are achieved without increase of density (even with a slight reduction of n$_{e,ave}$), as would be expected. This is all consistent with the top right plot, where $F$ is represented as a function of P$_{rad}$/P$_{ECRH}$: in that case, all groups of discharges, including low density seeded detachment, converge smoothly on a linear dependence to the radiated fraction. Regarding shear position (displayed in the bottom row), as anticipated in the handful of discharges analyzed in figure \ref{fig2b}, it reaches a maximum at $\rho \simeq 1$ for n$_{e,ave} \simeq 7.5-8 \cdot 10^{19} $m$^{-2}$, then decreases with density. Detached plasmas in which n$_e$ and P$_{ECRH}$ are reduced keep the shear position suggesting again that the contraction of the plasma is related to the detached condition and not directly to the density. In agreement with that, there is a nearly linear relation between $\rho_{shear}$ and the P$_{rad}$/P$_{ECRH}$ ratio. Instead, seeded discharges display an opposite behavior: $\rho_{shear}$ increases with detachment in them. The reason for this is unclear and will require a more detailed analysis. Finally, the dependence of electron temperature at the shear on shear position is represented in figure \ref{fig9}: as the shear moves inwards for n$_{e,ave} > 7.5 \cdot 10^{19} $m$^{-3}$, T$_{e,shear}$ increases. The same trend is observed for discharges with reduced P$_{ECRH}$, although with mostly lower T$_e$ values for the hollow points, as could be expected due to their lower heating power. Instead, in the case of seeded discharges, shear position moves radially outwards over detachment (as seen in Figure \ref{fig8}), resulting into lower T$_e$ values. In all these figures the light circles fall on top of the standard scenario points, indicating a general good agreement between manual and automatic analysis. \\

\section{Discussion}\label{Disc}

\begin{figure}
	\centering
	\includegraphics[width=0.5\linewidth]{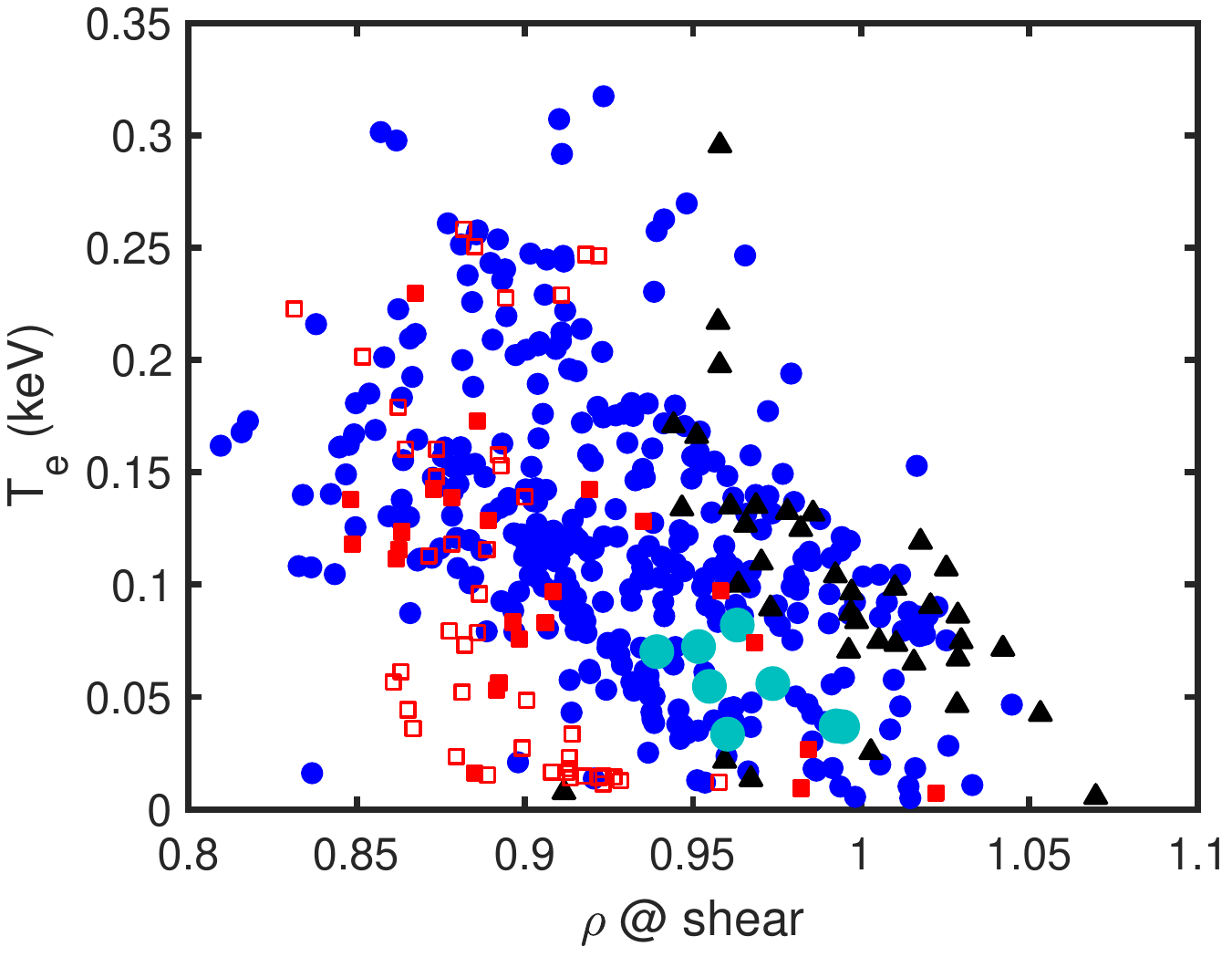}
	\caption{\textit{Relation between electron temperature and position of the shear. Symbols and colors as in figure \ref{fig8}.}}
	\label{fig9}
\end{figure}

Summarizing the results from previous section, detachment seems to cause a change in the shape in the plasma density profiles, which would become more peaked as P$_{rad}$/P$_{ECRH}$ increases. 
%A similar phenomenon was reported in W7-AS in equivalent discharges \cite{}.
The positive electric field measured in the SOL before detachment is very strongly reduced and seems to move into the region of closed magnetic surfaces, thus pushing the $\rho_{shear}$ to inner positions. If $\rho_{shear}$ indicates the position of the separatrix, this observation  -which is reproduced in a substantial number of discharges thanks to the automated analysis- would mean that, upon detachment, magnetic surfaces are deformed with respect to the equilibrium used to calculate $\rho$. However, there are at least two reasons against this interpretation: first, this would mean a displacement of the LCFS of several cm with respect to the vacuum magnetic configuration (in this case, a variation of $\Delta\rho = 0.1$ corresponds to 5 cm, approximately), which would require toroidal currents in the range of tens of kA \cite{Gao19}. Instead, as can be seen in figure \ref{fig5}, I$_{tor}$ is not correlated to the detachment (it reaches a maximum at $t = 4$ s and decreases afterward) and is always below $1$ kA. Similarly, the apparent displacement of the strike point which can be observed in figure \ref{fig5b} (approximately $1$ cm) is not correlated to the variation of toroidal current, but to the increase of $f_{rad}$. This indicates that it is not the result of a change in the magnetic configuration, but of the displacement of the heat flux peak after the detachment onset. Second, if the change of $\rho_{shear}$ were caused by a movement of the separatrix, it would be expected that T$_{shear}$ would stay roughly constant. Instead, as can be seen in figure \ref{fig9}, T$_{shear}$  increases at $\rho_{shear}$ up to values of T$_{shear} \simeq 200$ eV, which are substantially higher than those found at the separatrix before the detachment. Finally, since the same VMEC equilibrium is used for the whole discharge, such an apparent movement of the separatrix could be explained as the result of Shafranov shift if a substantial increase of $\beta$ was achieved during detachment. However, as can be seen in figure \ref{fig5}, that is not the case, as $W_{dia}$ remains essentially constant for the whole analyzed period.\\

If $\rho_{shear}$ is not indicating the position of the separatrix it follows that the ion-root negative electric field in the confined region is decreased substantially over detachment, eventually changing sign and becoming positive. While changes in E$_r$ at the SOL are to be expected due to the strong reduction in the T$_e$ in the target, it is not immediately clear how would detachment affect E$_r$ in the confined region. Since E$_r$ in the edge is determined by the ambipolarity equilibrium condition, this would require some mechanism altering charge conservation balance at the region. One such mechanism could be the resistive term opposing perpendicular current as a result of a a substantially enhanced ion-neutral collisionality in the edge. This could be caused by detachment as neutral density, $n_n$, increases as a result of the enhanced fueling caused by recycling at the wall and reduced $T_e$ at the plasma edge. The flux surface averaged radial current density resulting from this term \cite{Coronado93} would be 

\begin{equation}\label{eq5}
\left\langle \mathbf{j_n}\cdot\nabla r \right\rangle = m_i n_i n_n  (\langle v\sigma\rangle_{CX}+\langle v\sigma\rangle_{ion})\frac{1}{\left\langle B\right\rangle^2}\biggl(\frac{d\varphi}{dr}+\frac{1}{ne}\frac{dp}{dr}\biggr),
\end{equation}

with $\langle v\sigma\rangle_{CX}$ and $\langle v\sigma\rangle_{ion}$ being respectively the charge-exchange and ionization reaction rates and $p$ the ion pressure at a given flux surface. This mechanism cannot contribute to a positive E$_r$ in the edge, though, as it tends to oppose the existing electric field: as it can be seen in equation \ref{eq5}, $E_r > 0$ produces a positive current (the $E\times B$ term overcomes the diamagnetic one) which requires a reduction of E$_r$ in order to recover the ambipolar equilibrium. More importantly, the neutral densities required for $\left\langle \mathbf{j_n}\cdot\nabla r \right\rangle$ to be in range of kA/m$^{2}$ (the  order of magnitude of the neoclassical radial current density) would be unreasonably high, $n_n \gg n_i$. Therefore, the effect of neutral collisions is neither of the right sign nor magnitude to produce the observed results.\\

A better explanation is perhaps found in the substantial increase of ion collisionality, $\nu_i^*$, happening at the edge of the plasma as a result of the reduced temperatures, $\nu_i^* \propto T_i^{-2}$. For collisionality values beyond $\nu_i^*\gg 1$, neoclassical transport for ions takes place mostly in the Pfirsch-Schluter (PS) rather than in the plateau regime. In this asymptotic situation, the ion root radial electric field is roughly determined as \cite{Massberg99}:

\begin{equation}
 Z_i e E_r /T_i = n_i'/n_i - 0.5 T_i'/T_i, 
\end{equation}

where $n_i'$ and $T_i'$ are the radial derivatives with respect to $r$ and $Z_ie$ is the charge of the ions. Therefore, $E_r > 0$ if

\begin{equation}
|T_i'/T_i|>2|n_i'/n_i|. 
\end{equation}

Calculation of $\nu_i^*$ in the $\rho > 0.95$ region where E$_r > 0$ is measured by the DR is complicated by the fact that $T_i$ measurements by the XICS are not reliable so close to the LCFS and tend to overestimate it considerably. However, a qualitative discussion can be done given the low values of electron temperature achieved in this region after detachment ($T_e \simeq 50$ eV, as can be seen in figure \ref{fig7}), which make it reasonable to assume that both species are not far from thermalization and $T_i\simeq T_e$. In such case $\nu^*_i > 10$ is reached for $\rho > 0.9$, which would be more than enough to assume that most of the ion distribution is in the PS regime. As well, normalized gradients are typically higher for temperature than for density, specially in the case of electrons. Taking the data in figure \ref{fig7}, $|T_i'/T_i| \simeq 6 |n_i'/n_i| $ for $t = 2$ s, later rising to $|T_i'/T_i| \simeq 8.5 |n_i'/n_i|$ at detachment ($t = 7$ s). Therefore, it seems reasonable to expect that, the edge becomes more collisional due to radiation cooling and density increase, the neoclassical equilibrium E$_r$ starts to transit gradually towards values characteristic of the PS regime and eventually becomes positive at the outermost and most collisional part of the edge.\\

Finally, the observed displacement of the shear layer towards the plasma core could also be the result of an increase of the phase velocity of the turbulence due to either the increase of collisionality or to changes in the profiles. In this case, the initial hypothesis $v_E  \gg v_{ph}$ would no longer be valid, as both terms would become comparable. In this situation, changes in the the profile of $v_\perp$ measured by the DR could be dominated by the contribution of $v_{ph}$ and would not reflect changes in the E$_r$ profile. Linear simulations carried out at lower densities \cite{ref5} indicate that turbulence at $\rho \simeq 0.9-1$ is dominated by ITG modes. Assuming that this is still the case during detachment, $v_{ph}$ would point upwards at the measurement region (ion diamagnetic direction). Therefore, the observed displacement towards more positive E$_r$ values would indicate an increase of the phase velocity of the order of $5$ km/s. This would require a very substantial increase with respect to $v_{ph}$ values estimated in attached regimes: the aforementioned reference \cite{ref5} reports $v_{ph}\simeq 1-1.5$ km/s, while the agreement between experiment and simulation displayed in figure \ref{fig3} for the standard configuration suggests even lower values. This effect would be consistent with reports found in the literature in which phase velocity has been found to increase in regimes of high collisionality both in tokamaks \cite{Happel15} and stellarators \cite{Estrada19}. However, it must be taken into account that, unlike what is observed in W7-X, in those cases the increase of $v_{ph}$ was directed towards the electron diamagnetic direction.\\

Any of these two interpretations would be consistent with the gradual change observed in the E$_r$ profile as density increases and could also preliminarily explain the different behavior observed between density and seeding-induced detachment, as the second features both lower density and higher temperatures (as seen in figures \ref{fig8} and \ref{fig9}), thereby featuring substantially lower values of $\nu_i^*$ at $\rho_{shear}$. A detailed comparison of the two kinds of detachment is left for future work. As well, a proper calculation of the E$_r$ which includes this kind of effects is out of the scope of the present work and would require the use of neoclassical codes with a collisional operator which is valid in the PS regime, such as SFINCS \cite{Landreman14}, and reliable profiles of n$_e$, T$_e$ and T$_i$ for the outer edge and SOL. Similarly, an estimation of the phase velocity would require detailed simulations, eg. using gyrokinetic codes such as EUTERPE\cite{Jost01,Sanchez20} or Stella\cite{Barnes19}.\\

\section {Conclusions}\label{Conc}

As a general conclusion from this work, we can state that a first systematic survey of E$_r$ has been carried out using DR data, including the influence of configuration, n$_e$ and P$_{ECRH}$ on the radial electric field profiles. From this survey, a tendency of ion-root |E$_r$| in the edge ($0.7 < \rho < 0.9$) to increase with configurations featuring higher $\iota$ values has been observed, although the high mirror configuration does not follow it. This effect is partly, but not totally  due to flux compression: even taking $\left|\nabla r \right|$ effects into account, |E$_r$| is larger in high $\iota$ configuration than in the standard one for a similar density and heating power. As well, a dependency of |E$_r$| with power in this region has been observed for the standard configuration, EIM, which is not reproduced in the high mirror, KJM. The dependency of the shear amplitude and E$_r$ at the SOL has also been studied: $\Delta$E$_r$ and E$_{r,SOL}$ have been observed to decrease with n$_e$ in EIM, KJM and FSM configurations and to increase with P$_{ECRH}$ in EIM. This is consistent with current understanding of the relation between E$_{r,SOL}$ and the power flux onto the targets. As well, in good agreement with previous results, $\Delta$E$_r$ is larger in high $\iota$ configurations. Results have been compared to neoclassical simulations carried out with the codes DKES and KNOSOS, with good agreement between experiment and simulation in three of the four studied magnetic configurations (and qualitative agreement in the low iota case). In particular, the E$_r$ dependency on power (or lack thereof) has been reproduced correctly in EIM and KJM discharges. Finally, the evolution of profiles has been analyzed systematically through divertor detachment, including data from nearly $20$ discharges. The automatic analysis employed to estimate the shear position from DR data has been validated and is ready to be compared to other diagnostics. After detachment, |E$_r$| is reduced both at the SOL and edge. As well, the position of the shear moves inwards towards higher density and T$_e$ values. This is not the case when detachment is achieved by seeding, which causes the shear to move into the SOL instead. At the same time, the plasma column shrinks, as indicated by the form factor $F$ and the evolution of n$_e$ and T$_e$ profiles. This process is very well characterized by the radiated fraction of power, which reconciles impurity and hydrogen-induced detachment data, and accounts correctly for changes in P$_{ECRH}$ and density after detachment has occurred. The absence of strong enough currents in the plasma, indicate that magnetic surfaces are not significantly affected by detachment. Therefore, the observed strong reduction of $\rho_{shear}$ is probably the result of E$_r$ becoming positive inside the LCFS. This could be preliminarily explained as the result of the very high collisionality in the edge, either due to neoclassical transport transiting to the Pfirsch-Schlüter regime, or to a substantial increase of the phase velocity of the turbulence.

\section*{Acknowledgments}

The authors acknowledge the entire W-7X team for their support. This work has been partially funded by the Spanish Ministry of Science and Innovation under contract number FIS2017-88892-P and partially supported by grant ENE2015-70142-P, Ministerio de Economía y Competitividad, Spain and by grant PGC2018-095307-B-I00, Ministerio de Ciencia, Innovación y Universidades, Spain. This work has been sponsored in part by the Y2018/NMT [PROMETEO-CM] project of the Comunidad de Madrid. This work has been carried out within the framework of the EUROfusion Consortium and has received funding from the Euratom research and training programme 2014-2018 and 2019-2020 under grant agreement No 633053. The views and opinions expressed herein do not necessarily reflect those of the European Commission.\\

\end{document}